# Insect Diversity Estimation in Polarimetric Lidar

Dolores Bernenko[1*], Meng Li[1], Hampus Månefjord[1], Samuel Jansson[1], Anna Runemark[2], Carsten Kirkeby[3,4], Mikkel Brydegaard [1,2,4,5]

[1] Dept. Physics, Lund University, Sölvegatan 14c, 22363 Lund, Sweden

[2] Dept. Biology, Lund University, Sölvegatan 35, 22363 Lund, Sweden

[3] Dept. of Veterinary and Animal Sciences, University of Copenhagen, Copenhagen, Denmark.

[4] FaunaPhotonics, Støberigade 14, 2450 Copenhagen, Denmark

[5] Norsk Elektro Optikk, Østensjøveien 34, 0667 Oslo, Norway

* Corresponding author:

Email: dolores.bernenko@fysik.lu.se

# Abstract

Identification of insects in flight is a particular challenge for ecologists in several settings with no other method able to count and classify insects at the pace of entomological lidar. Thus, it can play a unique role as a non-intrusive diagnostic tool to assess insect biodiversity, inform planning, and evaluate mitigation efforts aimed at tackling declines in insect abundance and diversity. While species richness of co-existing insects could reach tens of thousands, to date, photonic sensors and lidars can differentiate roughly one hundred signal types. This taxonomic specificity or number of discernible signal types is currently limited by instrumentation and algorithm sophistication. In this study we report 32,533 observations of wild flying insects along a 500-meter transect. We report the benefits of lidar polarization bands for differentiating species and compare the performance of two unsupervised clustering algorithms, namely Hierarchical Cluster Analysis and Gaussian Mixture Model. We demonstrate that polarimetric properties could be partially predicted even with unpolarized light, thus polarimetric lidar bands provide only a minor improvement in specificity. Finally, we use physical properties of the clustered observation, such as wing beat frequency, daily activity patterns, and spatial distribution, to establish a lower bound for the number of species represented by the differentiated signal types.

**Data Availability:** The data that support the findings of this study are available from the corresponding author upon reasonable request.

**Funding:** This research work was sponsored by the European Research Council (ERC), under the European Union's Horizon 2020 research and innovation program (grant #850463, 'Bug-Flash'). In additional the FORMAS, Swedish Research Council (2018-01061).

**Competing interests:** The authors declare no conflict of interest.

# Introduction

Abundance and diversity of insects is in decline (1–4) especially in regions with industrialized agriculture (5). This loss of biomass and ecological functions can imply serious consequences for food chains in ecosystems (6) and pollination services of our crops (7). Rapid changes for conservation require rapid diagnostic tools to assess insect abundance and diversity. Photonic approaches (8) such as photonic sensors (9,10) and entomological lidars (11,12) have the potential to count and classify free-flying insects *in situ* continuously with close to no running costs. To date, entomological lidar can detect more than $10^5$ insects daily (13) and differentiate more than a dozen groups (11,12). While the count rate is superior to sweepnetting (14), traps (15) and robotic analysis (16), the taxonomic specificity is inferior to classification by e.g. machine vision (17) and genetic approaches (14). The non-intrusive nature of photonic approaches excludes post examination of the identified specimens. On the other hand, photonic *in situ* observations of insects provide complementary information which could not be obtained otherwise. For example, daily activity patterns (12), preferences for topographic features (18), or information on the species abundance distributions (19).

The number of insect species that can be identified by lidar or photonic sensors may be constrained by: a) the performance of the data clustering approach, b) the number of spectral (20,21) or polarization (22–24) bands of the instrument, or, in the ideal case, c) the number of present species in the habitat. The latter can reach more than tens of thousands co-existing species (25) out of the approximately ten million estimated insect species worldwide, amounting to an overall higher number of groups constituted by sexes, phenotypic variation, and appearance changing with the age of the specimens.

Most proposed photonic clustering of insects is based on assessing the wingbeat frequencies (WBF) (9,26). Insect WBFs range from approximately 10 Hz to 1000 Hz, however, the relative spread for a single species and sex under constant environmental conditions is generally 25%, which only leaves room for 18 distinct WBFs within this range. Wingbeat harmonics can provide additional information on wing dynamics (27) and specularity of the wings (28,29), thus improving specificity. Multiple studies have exploited wingbeat harmonics to differentiate insect groups (30). Even sexes from a single species can produce distinct harmonic content depending on observation aspect (22,24,31), with females generally being larger and having slower WBFs (32). WBFs are also influenced by temperature (32–34). However, in many cases, closely related species could produce similar signals indistinguishable for the instrument and setup. Nevertheless, species-rich insect ensembles will generally produce more diverse ensemble of signals (9).

Multiple studies have highlighted how multiple wavelengths could aid the differentiation of closely related species (22,28,35,36). In particular, specular flashes can be expected to be highly sensitive to the ratio of laser wavelength to wing membrane. Also, wing membrane thicknesses are frequently highly species-specific (28).

To what extent polarimetric information could improve specificity is less well-characterized. Generally, light loses its original polarization by multiple scattering in biological tissue (37). Consequently, near-infrared (NIR) light depolarizes when interacting with larger probe volumes in insect bodies on the scale of millimeters (22,31,38), whereas polarization is maintained when light probes thin insect wings on the order of a micron (28,39). Factors increasing the degree of linear polarization (DoLP) include absorption by melanin and water, which primarily punish photons with longer interaction path lengths that are more prone to depolarization. Factors reducing DoLP include wing scales of moth and butterflies (29) and even eggs inside the abdomen (40), which increase multiple scattering. However, it remains unknown to what extent polarimetrics could aid species differentiation.

In this paper, therefore, we investigate the benefits of polarimetric information for clustering of free-flying wild ensembles of insects. We report 32,533 insect polarimetric lidar observations, in a 500 m long transect over a lake. We use two unsupervised clustering methods to estimate signal diversity with and without polarimetric information. We attempt to assess to what extent diverse signals derive from a single species by analyzing the similarity of daily activity patterns and spatial distributions.

# Data Collection

## Field site

Field work was conducted on June 14th, 2020, at Stensoffa ecological field station, Sweden (55°41'44"N 13°26'50"E). The field site includes a forest, graze land, pond and a swamp (41), with low level of light pollution and high species richness. Within this site, we placed the experimental setup over a 500 m long, homogeneous, artificially created peat pond. A Scheimpflug lidar was positioned on one shore with the termination point on the opposing shore. Both the lidar and termination point maintained a constant height over the pond throughout the transect, with mostly the same distance to the shore on both sides of the transect.

By selecting a rectangular pond, we aimed to minimize the influence of topological differences on insects flying across the laser beam, for example, due to differences in vegetation or flight distance between shores. However, some parts of the beam were visited by insects more frequently due to the presence of patches of reeds and floating water plants.

## Instrument

The design of the Scheimpflug lidar system is described in (42). It is based on kHz time-multiplexing, comprising two TE polarized 3W, 980 nm laser diodes (MLD-980-3000, CNI lasers, China). The laser apertures are 95μm and fast-axis-collimators (FACs) are glued to diodes reducing their divergence to 8° in both axes. A NIR wavelength was chosen to avoid disturbing the insects, as they are insensitive to this light. Furthermore, backscattering is increased at this wavelength because insect melanization absorbs less NIR light.

To retrieve polarimetric lidar data, we illuminate the targets with laser beams of alternating orthogonal linear polarization. To achieve this, we rotate the polarization state of one of the laser sources by 90° using a half-wave plate (WPQ10E-980, Thorlabs, USA), then co-align the two beams using a polarizing cube beam splitter (PBS203+B4CRP/M, Thorlabs, USA). The radiation is collimated by a Ø75 mm, f = 300 mm achromatic doublet (#88-597, Edmund Optics, UK) in a focus mechanism (Monorail, Teleskop-Service, Germany). The lidar overlap is controlled by a tangential mount (Stronghold, Baader planetarium, Germany). The receiving telescope is a Ø200 mm, f = 800 mm Newton reflector (Quattro, SkyWatcher, China). The received light passes a 10nm FWHM filter at 980 nm (#65-247, Edmund Optics, UK) and a NIR linear polarizer (LPNIRE200-B, Thorlabs, USA) before it is imaged onto a linear CMOS detector, which is tilted 45° according to the Scheimpflug condition and hinge rule. The linear array detector (OctoPlus, Teledyne e2v, USA) has 2048 pixels of 10x200 μm each. It can read out 80 kLines/s at 12 bits, but in this experiment, it was operated at 6 kHz.

Our system achieves kHz-rate separation of co-polarized and de-polarized light components by multiplexing two orthogonal laser sources (43,44). We sequentially illuminate the target with a three-timeslot cycle: timeslot 1, laser I is ON; timeslot 2, laser II is ON; timeslot 3, both lasers are OFF (used for real time subtraction of the background from the first two exposures). This effectively provides a 2 kHz sample rate with a maximum observable modulation frequency of 1 kHz due to the Nyquist criterion (45). The lowest achievable frequency and resolution depend on the insect's transit time through the laser beam.

## Lidar observations

We conducted continuous lidar recordings throughout June 14, 2020, accumulating ~2.5 terabytes of raw data. To isolate insect observations, we implemented a thresholding technique, selecting data exceeding the median intensity of backscattered light plus five times the interquartile range (IQR) within each 5-second data file (~30,000 exposures), see (13,46,47) for detailed accounts of the preprocessing. We further refined the dataset to include only observations exceeding 40 ms transit time, corresponding to a minimum detectable WBF of 25 Hz. This criterion yielded a total of 32,533 observations. A typical insect observation manifests as a modulation of backscattered light intensity over both time (exposure number) and space (pixel number), as illustrated in **Fig 1a**.

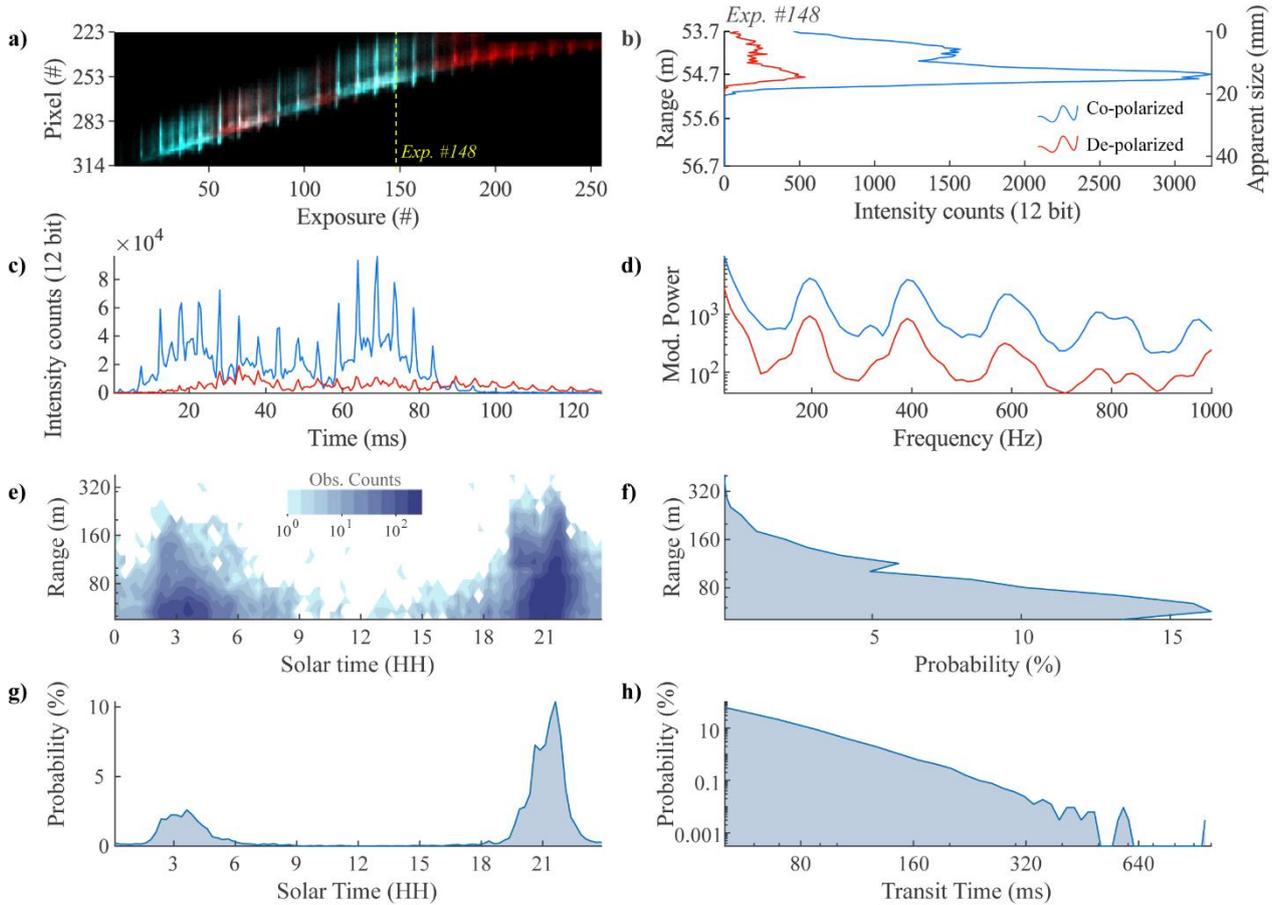

**Fig 1. Lidar insect observations**. (a) Modulation of backscattered light intensity from a single insect across exposures (time domain) and pixels (space domain). Co-polarized (cyan) and de-polarized (magenta) components shown. (b) Instantaneous echo in the range domain (@ exposure #148), with range and insect size deduced from absolute and differential pixels respectively. (c) Signal waveform showing intensity modulation over time. (d) Power spectra. (e) Distribution of observations by solar time (15-minute bins with bin centers from 00:07 and ending at 23:53) and range (20 logarithmically spaced bins between 48 m and 427 m). Time is reported in true solar time. (f) Range distribution of insect observations. (g) Time distribution of insect observations. (h) Distribution of insects' transit times >40 ms. In (b-d), co-polarized components are in red, de-polarized in blue, see legend in (b).

We analyzed the lidar signal in several ways. First, projecting the signal into the spatial domain provides lidar echo intensity across pixels. This information can be used in two ways: 1) by transforming absolute pixel numbers to determine the distance to a target (left y-axis in **Fig 1b**), and 2) by transforming differential pixel numbers to estimate the apparent insect size (right y-axis in **Fig 1b**).

Second, analyzing the signal from the co-polarized and de-polarized channels in the time domain generates two waveforms (**Fig 1c**). Comparing these waveforms, we observe that co-polarized backscatter from glossy wings manifests as a series of brief, specular flashes. In contrast, the de-polarized backscatter lacks these distinct flashes and instead presents less intense, smoother waveform with the same periodicity, caused by broader scattering lobes by the de-polarizing wing features such as the veins and scales. The relative intensities of co-polarized and de-polarized light are also informative. For example, nearly equal intensity in the co-polarized and de-polarized waveforms suggests that most of the backscattered light has a randomized polarization state (thus an equal chance to detect co-pol. and de-pol. signals), while a dominant co-polarized signal indicates a higher degree of glossiness.

We also explored the temporal and spatial distributions of the observations. **Fig 1e** visualizes a 2D histogram illustrating the count distribution, while **Figs 1f** and **1g** show the probability of observations based on range and solar time. Notably, few observations are detected close to noon, and it is more likely to detect an insect closer to the detector. Additionally, we present a transit time histogram (**Fig 1h**) displaying the distribution of transit times for all observations exceeding the 40 ms threshold.

By combining spatial, temporal, and polarimetric information, we can characterize each insect observation and identify broader patterns within groups. For example, in the waveforms, periodic bright reflections correspond to the insect's WBF, while the duration of these flashes can indicate wing specularity. By comparing the intensity of co-polarized and de-polarized backscatter, we can quantify the DoLP. This combined analysis allows us to differentiate insects with similar WBF but distinct polarization signatures. Additionally, we can determine the detection range and time of day for each observation, or analyze these distributions for a group, revealing time activity patterns and spatial preferences for groups of insects.

## Estimation of oscillatory power spectra

Despite waveforms being highly informative, directly comparing them for insect clustering is challenging. Variations in waveform shape can arise from external factors, such as the insect's time spent within the lidar beam, and the independent phases of wingbeat and lidar sampling. To address this, we calculate the oscillatory power spectra for each observation (**Fig 1d**), which represent the signal in frequency domain as a distribution of power across normalized frequency bins. The resulting power spectra reveal the insect's fundamental WBF and its harmonic overtones, providing a more robust basis for clustering and comparison.

To estimate the power spectral density, we use Welch's method, implemented in MATLAB Signal Processing Toolbox. We define the observable frequency range spanning between 25 Hz (reciprocal of minimal transit time) and 1000 Hz (the Nyquist frequency), and the number of linearly spaced frequency bins as 80 (the number of time samples in 40 ms-long observation at 2000 Hz sampling frequency). We also define a Gaussian time window with a FWHM of half the number of time samples. We set the number of overlapping samples in the sliding Welch power estimate to 79, the maximum possible overlap constituting the heaviest computations operation.

## Power spectra preprocessing

While power spectra capture an insect's wingbeats in a fundamental peak and wing glossiness as the number of harmonic overtones, we hypothesize that incorporating polarimetric data may reveal additional distinctions based on wings' DoLP. To test this hypothesis, we generated three datasets representing different data acquisition scenarios: with and without polarimetric data.

**Non-polarimetric data acquisition (unpolarized dataset)**

This dataset simulates a scenario when a signal is acquired without polarimetry. We achieve this by summing up both co- and de-polarized power spectra and then normalizing the area under the merged curve to unity (**Eq. 1**).

$$P_{unpol}(f) = \frac{P_{co}(f) + P_{de}(f)}{\sum[P_{co}(f) + P_{de}(f)]} \qquad (1)$$

Here, $P_{unpol}(f)$ is the unpolarized power spectrum, $P_{co}(f)$ and $P_{de}(f)$ are the co-polarized and de-polarized power spectra, respectively.

We show the resulting power spectrum in **Figs 2a** (specular case) and **2d** (diffuse case). By color-coding the proportion of the de-polarized signal, we illustrate the similarity between the unpolarized (total) signal and the de-polarized signal. We observe that in a specular case, de-polarized light improves the certainty of the peak at ~250 Hz, however, and has little influence on other frequency peaks. Whereas, in diffuse case, de-polarized light is the main contribution to powers.

**Coherently backscattered light acquisition (co-pol. dataset)**

To obtain the co-polarized dataset, we take only the co-polarized component and normalize it to unity (**Eq. 2**). This represents an acquisition scenario, when targets are illuminated using linearly polarized light, and measurements made in the same polarization state (**Figs 2b, 2e**).

$$P_{co}^*(f) = \frac{P_{co}(f)}{\sum P_{co}(f)} \qquad (2)$$

**Polarimetric data acquisition with Degree of Linear polarization (DoLP dataset)**

The DoLP dataset (**Figs 2c, 2f**) is a scaled version of the co-polarized dataset. In this dataset, the area under the co-polarized power spectrum represents the DoLP information for the oscillatory part of the signal, excluding the 0-25 Hz range (**Eq. 3**).

$$P_{DoLP}(f) = \frac{P_{co}(f)}{\sum [P_{co}(f) + P_{de}(f)]} \qquad (3)$$

Importantly, when normalizing the areas under all power spectra, we ensured that the relative strength of frequency components within each spectrum remains consistent regardless of the distance at which the insect was observed. This approach addresses a potential source of bias in our analysis—namely, the signal intensity attenuation with distance.

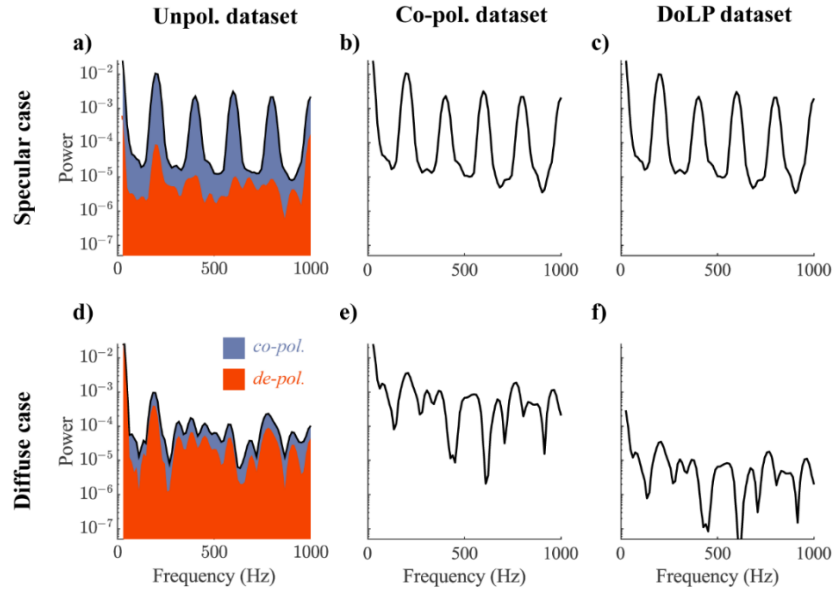

**Fig 2. Three datasets with varying polarimetric information for a specular (top row) and a diffuse observation (bottom row)** (a, d) Unpolarized data is shown as black solid line, whereas blue shade shows contribution from the co-polarized channel, and orange – from de-polarized; (b, e) Co-polarized dataset. (c, f) DoLP dataset.

# Results and discussion

## Cluster count and agreement analysis: HCA vs. GMM

Unsupervised clustering is a valuable tool for rapidly assessing insect diversity from lidar observations. Unlike classification, which requires labeled data that is often scarce and costly to obtain, clustering groups insect observations based on inherent similarities in their characteristics. This study focuses on characteristics embedded into power spectra of observations, specifically the frequency content (WBF and harmonic overtones) and the DoLP (when using the DoLP dataset).

However, these features may not sufficiently distinguish among insect species, as WBF can be common across multiple species and exhibit significant variability even within the same species. This feature overlap can cause multiple species to merge into clusters or a single species to split into multiple clusters, affecting our conclusions on insect diversity estimates. Additionally, diversity estimates could be biased due to different clustering algorithms producing different solutions, that vary in the number and size of identified clusters.

In this section, we explore the differences between clustering solutions by employing two contrasting methods. One is Hierarchical Clustering Analysis (HCA), a deterministic approach previously employed to group observations from photonic sensors and lidar (9,11,12,19) (see **Methods: HCA**), and Gaussian Mixture Model (GMM), a stochastic approach (see **Methods: GMM**). Comparing HCA and GMM clustering results, we observed that these methods clustered lidar observations with varying granularity. HCA yielded 803 (unpolarized), 245 (co-polarized), and 256 (DoLP) clusters, while GMM produced fewer: 80 (unpolarized), 86 (co-polarized), and 89 (DoLP).

To determine if these methods produce consistent results despite the varying granularity, with HCA offering a more fine-grained view, we assessed the agreement between their clusterings using two metrics: Adjusted Mutual Information (AMI) and Homogeneity score. AMI ranges from 0 to 1, with higher values indicating that the same observations are grouped into the same clusters across both methods, after adjusting for chance. The Homogeneity score, also ranging from 0 to 1, evaluates whether each cluster from one method contains observations primary from a single cluster in the other method. A high Homogeneity score indicates that one method's clusters are subsets of the other's. We explain both metrics in detail in Section **Methods: Evaluating clustering agreement**.

We observed moderate agreement between the methods, with AMI scores ranging from 0.47 to 0.55 (**S1a Fig**). However, the Homogeneity score was higher, ranging from 0.66 to 0.74 (**S1b Fig, the upper triangle**). This result suggests that there is a difference in the underlying composition of clusters, and that the methods did not merely identify the same clusters at different resolutions. Despite these differences in the number and composition of clusters, most clusters in both solutions exhibited discernible frequency content (see median power spectra for clusters in **S2-S3 Figs**). In the absence of ground truth for optimal partitioning, we then evaluated clustering results based on DoLP homogeneity, distinction in activity time patterns, and spatial distribution.

# Degree of linear polarization for clusters

In this section, we investigate whether wings' polarimetric characteristics (from glossy to diffuse) can be predicted using unpolarized data alone, and how this prediction is improved by including polarimetric data. To quantify the differences between datasets, we measure the clusters' DoLP homogeneity as detailed in **Methods: Bootstrapping to evaluate confidence intervals.** We report the clusters' homogeneity as the mean DoLP and its 95% confidence interval (CI) ($2.5^{th}$ and $97.5^{th}$ percentiles). To determine the significance of the observed results, we compared the CIs of a mean DoLP for found clusters against those derived from randomly assembled clusters of the same size. We also divided clusters into four groups based on DoLP quartiles (from $Q_1$, most glossy, to $Q_4$, most diffuse).

We find that most of clusters from the glossy group ($Q_1$) and some from the diffuse group ($Q_4$) are significantly different from random ones (CIs of found and random clusters do not overlap), see **Fig 3** (DoLP dataset) and **S4-S5 Figs** (unpol. and co-pol. datasets). The clusters' DoLP uncertainty is largest for the HCA applied to the unpolarized dataset (**S4a Fig**), however, this dataset returns smaller clusters. The major difference between the three datasets is that including polarimetric information improves isolation of low-DoLP observations into distinct clusters. Notably, HCA shows greater sensitivity in finding clusters with lower DoLP compared to GMM. Intriguingly, both methods identified clusters with anomalously low DoLP (~1-2%), suggesting a less than random polarization state for the backscattered light. Potential explanations include scattering from extremely small, fluffy insects where polarized light escapes on the backside before having the chance to scatter 180°. It could also be measurement outliers due to imperfect beam overlap.

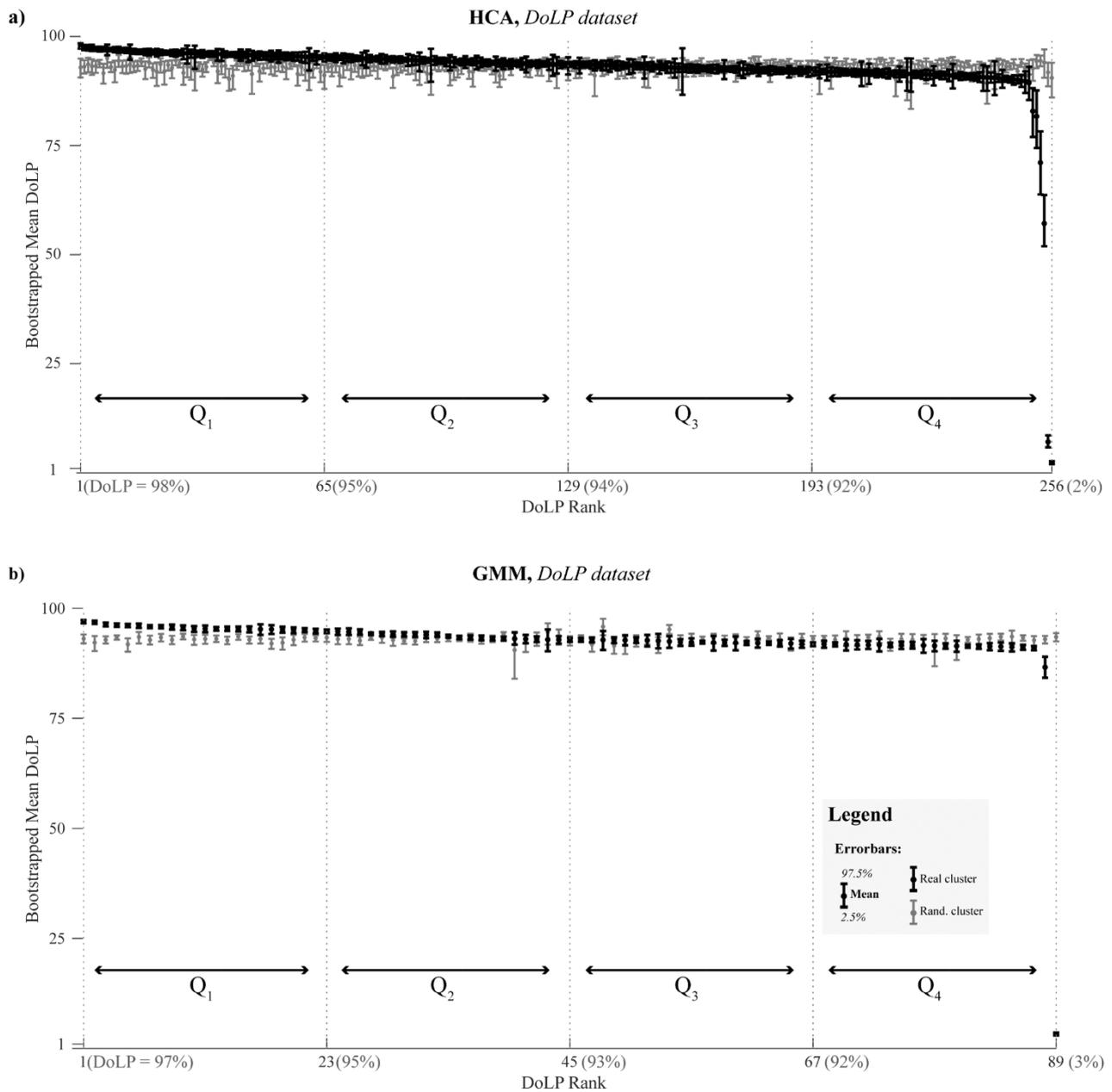

**Fig 3. DoLP characterization of clustering results.** (a) HCA and (b) GMM show comparisons of cluster DoLP distributions for found clusters (black error bars) and randomly generated clusters of the same size (gray error bars). Error bars represent the bootstrapped mean DoLP and its 95% CI for each cluster. Found clusters are ranked by decreasing mean DoLP (x-axis). Vertical lines denote DoLP quartile boundaries ($Q_1$-$Q_4$).

To further examine the impact of polarimetric information on clustering results, we visualized the rearrangement of observations across different DoLP quartiles **(Fig 4)**. We aggregated observations based on the DoLP of their assigned cluster and represented these rearrangements using flow lines. Our analysis shows that the $Q_1$ quartile produces the most consistent results, with 26% of $Q_1$ observations being shared across the three datasets in HCA and 37% in GMM. Significant rearrangements between unpol. and DoLP datasets predominantly occur between adjacent quartiles, though 9% (HCA) or 12% (GMM) observations are reassigned across non-adjacent quartiles (e.g., from $Q_1$ to $Q_4$). We conclude that even without polarimetric information, clustering algorithms can identify highly glossy wings. However, polarimetric data is particularly beneficial for co-clustering together low-DoLP observations.

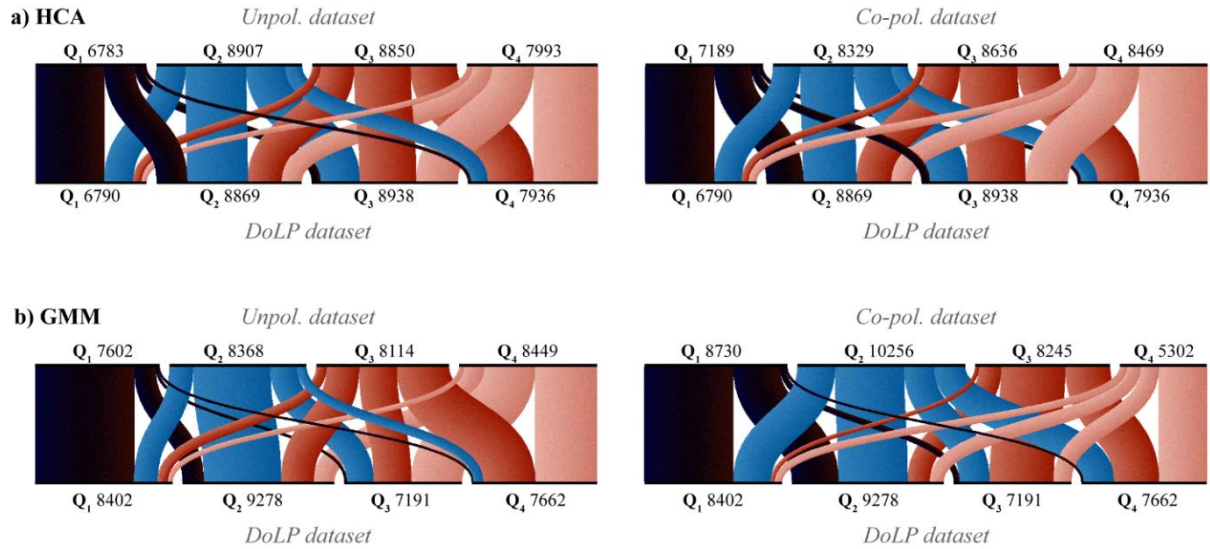

**Fig 4. Rearrangement of observations between cluster's DoLP quartiles**. (a) HCA clusters. (b) GMM clusters. Left panels show rearrangements between the unpolarized and the DoLP datasets, and right panels illustrate differences between co-polarized and DoLP datasets. Each quartile ($Q_1$–$Q_4$) is labeled with the number of observations. The flows (lines) between quartiles indicate the fraction of observations, with line width proportional to the number of observations. To plot the alluvial diagrams we use RAWGraphs (48).

To evaluate if HCA and GMM agree on the content of the top five glossy clusters, we next compare their median power spectra (DoLP dataset, **S6-S7 Figs**). Despite both returning similar power spectra for rank 1 and 2 clusters, GMM aggregates more observations per cluster (e.g., rank 1: 123 observations in GMM vs. 23 in HCA). This indicates that GMM generalizes power spectral patterns more broadly, leading to larger clusters, while HCA maintains a stricter similarity criterion. The conclusion is thus the same when based on similarity of top two glossy clusters as when based on the homogeneity score.

## Time and range communities

Distinct species are likely to exploit distinct niches in time and space. This could be a matter of crepuscular species adapted to a certain ambient light level or bumble bees adapted to forage earlier in the colder mornings. In terms of range, preferences for topographic features such as vegetation or reeds along the transect could differ. It could also be biased by the resolution of the instrument since larger, brighter, or glossier species could be detected over further ranges.

To further assess the biological relevance of clustering, we investigated whether distinct daily activity patterns and range profiles could define communities – groups of clusters that are more similar within a group than between (see Section **Methods: Time and range communities**). Comparing two clustering approaches, we find that GMM method most clearly recovers community structure, whereas HCA performs worse. We quantify it using a modularity metric (M). It ranges from 0 (random structure), to 1 (well-defined structure), or to -1 (less optimal than random). In HCA, modularity increased with the addition of polarimetric information (unpol. < co-pol. < DoLP). This trend was evident in both time communities ($M_{unpol.}$ = 0.07, $M_{co-pol.}$ = 0.15, $M_{DoLP}$ = 0.16) and range communities ($M_{unpol.}$ = 0.08, $M_{co-pol.}$ = 0.13, $M_{DoLP}$ = 0.14). In contrast, clusters identified by GMM show relatively strong community structure across all datasets, with modularity remaining consistent for both time ($M_{unpol.}$ = 0.26, $M_{co-pol.}$ = 0.27, $M_{DoLP}$ = 0.25) and range communities ($M_{unpol.}$ = 0.12, $M_{co-pol.}$ = 0.09, $M_{DoLP}$ = 0.11). The presence of community structure indicates that the time and range profiles of the clusters diverge from the average pattern, suggesting ecologically distinct groups. However, the moderate modularity scores imply these patterns are not discrete but rather overlapping, with some clusters exhibiting similarity to multiple communities. This is visualized in **Fig 5**, a heatmap of cluster-to-cluster similarity, where communities appear as bright squares along the diagonal, but some clusters show high similarity across communities.

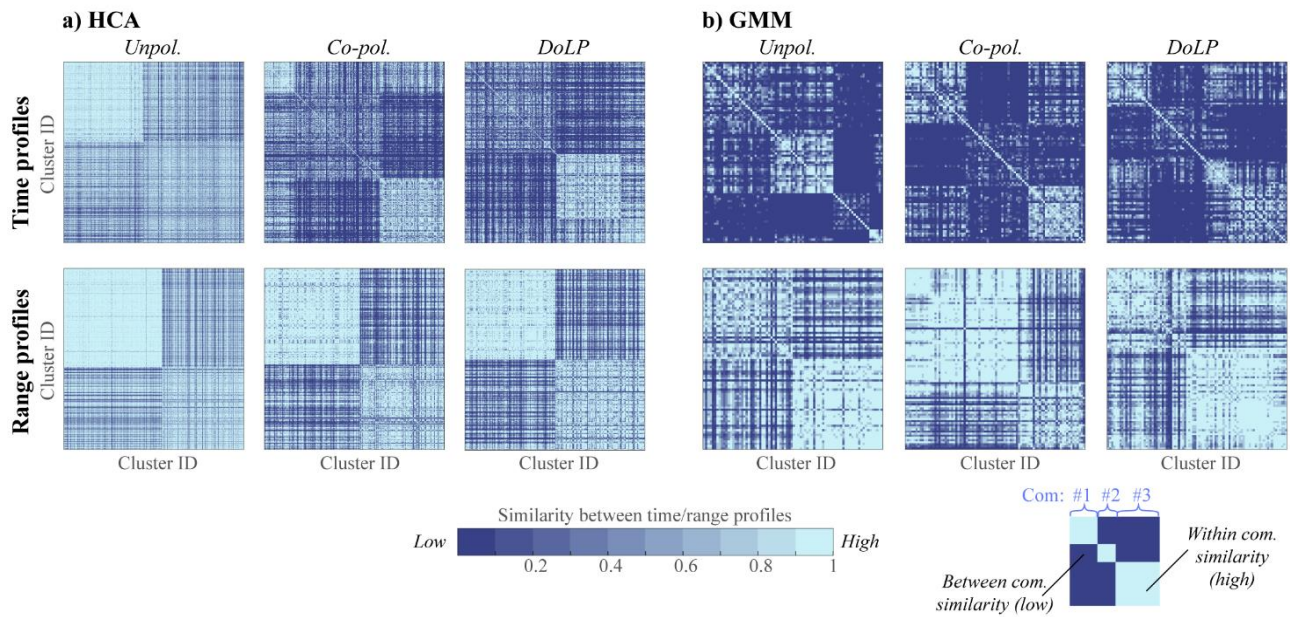

**Fig 5. Community structure analysis for HCA and GMM clustering results based on time and range profiles.** Each symmetric matrix displays similarity of time (top panels) and range (bottom panels) profiles across cluster pairs, with darker pixels indicating greater dissimilarity. The heatmaps are organized to place rows/columns adjacently if clusters are from the same community, thus making communities to appear as bright squares along the matrix diagonal manifesting greater similarity within a community then between them (see the bottom-right schematic).

Next, we characterized both time and range communities by plotting communities' probability distributions across time and range bins (**Figs 6** and **S8**). To illustrate clusters' variability, we employed bootstrapping (see **Methods: Bootstrapping to evaluate confidence intervals**). We observe that time communities primary differentiate based on variation in evening activity patterns (**Fig 6 I-III**), whereas range communities are characterized by a decaying probability of an observation with a different detectability cut-off: with some clusters detected at mid-ranges, <160 m (**Fig 6A**) and others primarily at long ranges, <255 m (**Fig 6B**).

We hypothesize, the variation in spatiotemporal profiles may be related to the frequency content of the lidar signal. To visualize this, we plot clusters' median power spectra after detrending (see **Methods: Detrending of power spectra**) showing them as heatmaps at the intersection of (A, B) and (I-II-III) probability plots (**Fig 6B**). Here, we observe that insects detected at long ranges (group B) tend to have a first peak in their frequency spectrum below 250 Hz. This peak could correspond to the fundamental frequency of a wingbeat, suggesting that larger insects, which have lower WBFs, are more likely to be detected at greater distances (for example, predating dragonflies with up to 14 cm wingspan).

## Range dependence of co-polarized backscatter

To further explore the factors influencing long-range detectability, we investigated the impact of wing glossiness. We hypothesize that insects with glossy and clear wings scatter laser light coherently, with a narrow lobe and rapid angular speeds, resulting in improved transmission over distances. To test this hypothesis, we subdivided insects from the range communities (A: mid; B: far) into four quartiles based on their DoLP ($Q_1$-$Q_4$, representing decreasing glossiness, see **Fig 3**). Creating these subsets of clusters allows us to compare range profiles of, for example, highly glossy insects detectable at far ranges ($Q_1$-B subset of clusters) with diffusive insects detectable at the mid-range ($Q_4$-A). Next, for each subset, we calculated the mean probability of detection at each range bin, along with the 2.5[th] and 97.5[th] percentiles (CIs), as described in section **Methods: Bootstrapping to evaluate confidence intervals**.

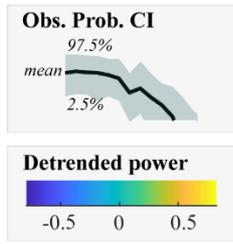
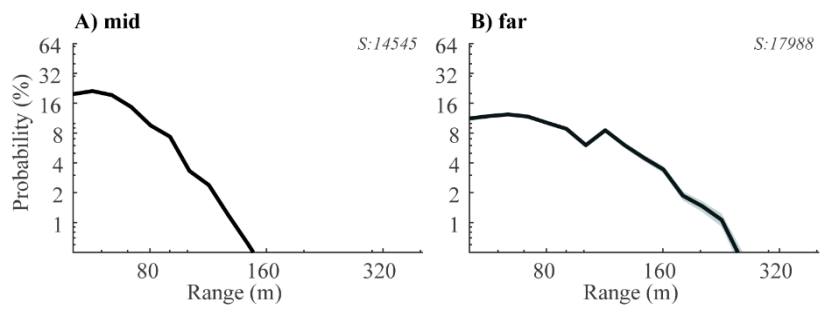
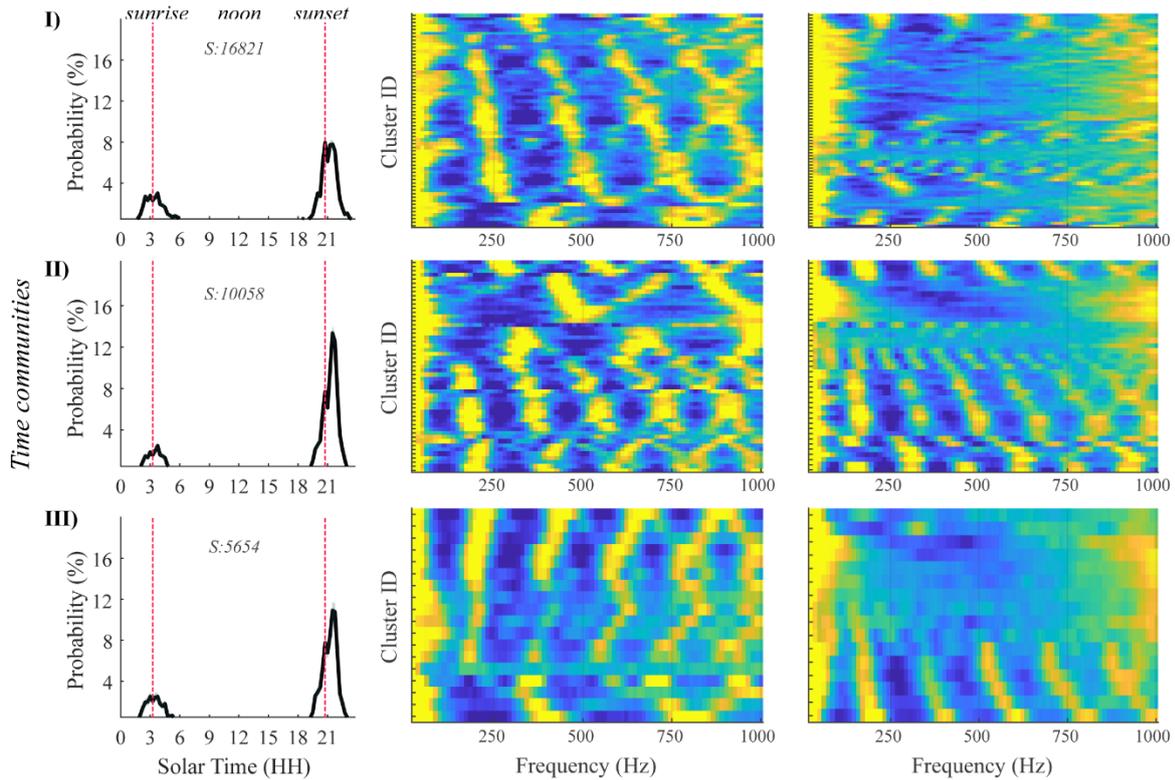

**Fig 6. Characterization of time and range communities.** CIs of observation probability for (A, B) range and (I-II-III) time communities (see legend). At the top of each panel, we show the size of a community. Heatmaps at the AB and I-III intersection display median power spectra for a corresponding time-range community. The y-axis segments heatmaps into stripes, one for each cluster. Variation of colors within a stipe indicates power magnitude at corresponding frequencies (x-axis). The powers are shown after normalization, logarithmic transformation, and detrending. The color-bar encompasses 5th to 95th percentiles of all range of power values.

Comparing range profiles for different DoLP groups, we observe a striking feature in the far-range community: a peak at ~120m in an otherwise decaying with distance probability of observation (**Fig 7** and **S9 Fig**). This peak is most prominent for glossy insects ($Q_2$). Visualizing the laser beam path over the pond (**Fig 7**, bottom), we note that this peak coincides with the proximity of a landmass, marked with a red dot. This suggests differences in insect communities based on proximity to land. Acknowledging the noise introduced by assuming that observations from all DoLP groups ($Q_1$-$Q_4$) have an equal probability of being present at this landmass, we hypothesize that the lack of a peak at 120 m in the low-DoLP distributions (particularly $Q_4$) implies that glossiness significantly affects detectability at this distance.

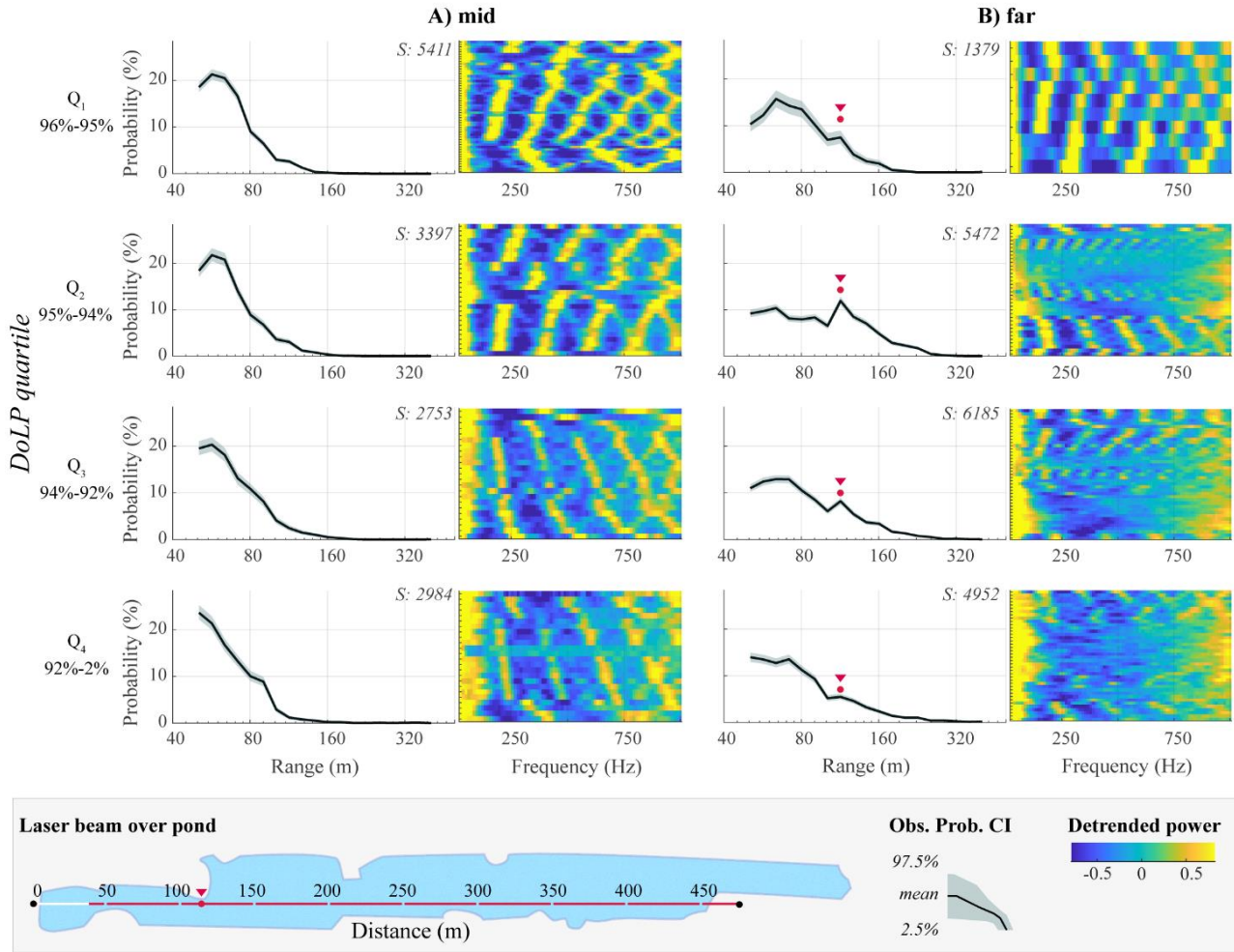

**Fig 7. Range dependence of co-polarized backscatter.** CIs of probability distributions show the likelihood of observing an insect within a DoLP quartile ($Q_1$-$Q_4$: glossy to diffuse) and range community (A: mid-range, B: far-range). In the top-right corner of probability distributions, we show the number of observations. In B-plots, we show with the red dot the spike in the probability of observing an insect, potentially linked to a nearby landmass (see bottom panel). Heatmaps depict median power spectra for clusters within corresponding DoLP-range subsets (as in **Fig 6**).

These findings indicate that the clusters reflect spatial preferences of insects and thus could be seen as a meaningful coarse-grained representation of lidar observations. This representation can be further employed to describe insects' activity patterns and spatial preferences, for example, due to changes in vegetation over seasons, or to provide a means for evaluating the attraction range of conventional insect traps.

Our findings also highlight some limitations of the current lidar setup in assessing biodiversity. Specifically, there are biases in determining the abundance and richness of insects. For example, some morphological features make certain insects easier to detect, leading to overestimation of their presence. These features could be size, brightness, and glossiness, and depend on how wing thickness resonates with the lidar wavelength. This observation suggests a direction for improving lidar technology by using longer wavelengths to enhance specularity and detection range. Longer (infrared) wavelength have proven efficient in clustering moths (29,49).

# Lidar based diversity indices

We hypothesized that integrating polarimetric information into lidar signals would enhance discrimination between insect taxa, leading to a rearrangement of observations into clusters based on both the frequency content of power spectra and the similarity of polarimetric properties of insect wings and bodies towards low frequencies. However, clusters' count and composition depend not only on the instrument but also on the choice of clustering algorithm, influencing conclusions about the diversity at the monitored site. To evaluate the impact of clustering approaches on diversity estimates, we compared the results of HCA and GMM clustering, focusing on the number and relative size of the identified clusters.

To illustrate cluster count and their relative size, we plotted the Ranked Abundance Distribution (RAD), depicting cluster sizes in descending order (**Fig 8**). We further characterized clustering results using Hill numbers, a family of diversity metrics (see **Methods: Lidar based diversity indices**). Specifically, $H_0$ represents the total cluster count, providing an overall estimate of diversity; $H_1$ represents the effective number of clusters, accounting for relative abundance; and $H_2$ represents the dominant number of clusters, highlighting the most prevalent clusters.

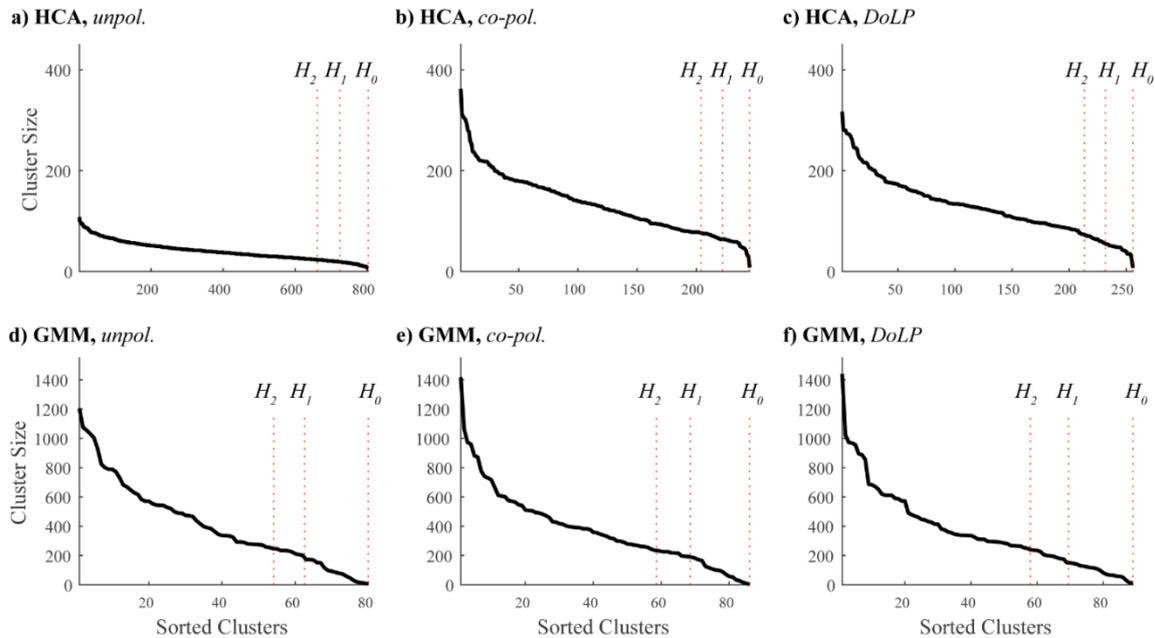

**Fig 8. Clusters' size distribution.** (a–c) HCA clustering on three datasets; (d–f) GMM clustering on three datasets. The solid line shows the number of observations per cluster for clusters sorted from largest to smallest. Vertical lines mark Hill numbers.

Our analysis revealed a consistent trend of HCA producing a higher number of clusters compared to GMM (~250 vs. ~85), particularly evident in the unpolarized dataset (~800 vs. ~80) as illustrated in **Fig 8** and **Table 1**. This suggests that HCA clusters are generally more diverse than GMM clusters. However, the high homogeneity score (~0.7, **S1b Fig**) between the two clustering solutions indicates that larger GMM clusters are often fragmented into smaller HCA clusters. Thus, the higher diversity estimates obtained through HCA likely reflect a finer resolution level at which the data is partitioned.

To address the potential disproportionate influence of rare clusters on cluster richness ($H_0$), we further evaluated the cluster size distribution using the effective number of clusters ($H_1$). HCA consistently yielded a larger effective number of clusters than GMM relative to the total number of clusters, suggesting a more balanced distribution of cluster sizes. Moreover, HCA identified a substantially larger proportion of dominant clusters ($H_2$) compared to GMM (~90% vs. ~65%) (**Fig 8**), indicating that our diversity estimates were not significantly inflated by rare clusters.

**Table 1. Characterization of clustering results with Hill numbers.** NoC is a number of clusters.

|  | Dataset | $H_0$ (NoC) | $H'$ (Shannon Index) | $H_1$ (Effective NoC) | $H_2$ (Dominant NoC) |
|---|---|---|---|---|---|
| **HCA** | unpol. | 803 | 6.58 | 724 | 662 |
|  | co-pol. | 245 | 5.40 | 222 | 204 |
|  | DoLP | 256 | 5.45 | 232 | 213 |
| **GMM** | unpol. | 80 | 4.14 | 63 | 54 |
|  | co-pol. | 86 | 4.23 | 68 | 59 |
|  | DoLP | 89 | 4.24 | 69 | 58 |

Hill numbers reveal that each method can lead to distinct conclusions, particularly regarding the proportion of dominant clusters within the total cluster count. These discrepancies are largely due to HCA and GMM exhibiting different levels of tolerance for variation within clusters. HCA favors similarly sized, spherical clusters because of the Ward linkage criterion, which defines a "good" cluster as one where all observations are relatively close to the cluster centroid. In contrast, GMM identifies clusters based on the probability of an observation belonging to a specific Gaussian distribution, allowing for the identification of elliptical clusters. Consequently, these differences impact the number and size distribution of clusters, and subsequently, the estimated diversity indices. Therefore, when interpreting insect diversity estimates derived from lidar data, it's crucial to carefully consider the inherent biases and assumptions of different clustering algorithms.

To move beyond the limitations of single clustering solutions and ensure more robust lidar-based diversity assessments, future research would benefit from evaluating the robustness of these indices through a more comprehensive approach. One promising avenue involves using stochastic algorithms to analyze an ensemble of clustering solutions, rather than relying on a single outcome (50,51). This would allow us to report a range of values for each Hill number, gaining valuable insights into the sensitivity of these metrics in detecting changes within the monitored site (see additional analysis for GMM results in **S1 Text**). Additionally, focusing on observations that consistently co-cluster together across multiple solutions could provide a more reliable basis for diversity estimates, as these observations represent a stronger signal compared to those that are grouped inconsistently and may introduce unpredictable variability.

# Conclusions

Estimating insect diversity has traditionally been labor-intensive, relying on manual capture and classification (52). However, researchers have sought to automate this process (53) using technologies like radar (54) and lidar. In this study, we use polarimetric lidar to detect free-flying insects and investigate whether polarimetry improves diversity estimates. We hypothesized that diversity estimates would vary depending on the amount of polarimetric information included in lidar observations.

We initially focused on the total cluster count produced by each clustering method. We observed a distinct difference in resolution, with GMM yielding ~85 clusters and HCA ~250. However, when interpreting this value as an estimate of insect diversity, it's important to recognize that neither algorithm intrinsically determines the optimal number of clusters. In HCA, increasing the significance threshold for compensated linkage would lower the cluster count, while in GMM, minimizing the AIC instead of the BIC would increase it, yielding ~300 clusters per dataset. Therefore, this value should be seen as a lidar-based diversity index rather than a direct measure of insect diversity.

Regardless of clustering resolution, we aimed to determine which lidar signal (unpolarized, co-polarized, or DoLP) results in greater diversity estimates when comparing results within the same clustering approach. This analysis yielded conflicting results. GMM yielded fewer clusters for the unpolarized dataset than for DoLP (80 vs. 89), while HCA produced a significantly higher number (803 vs. 256).

To investigate whether HCA's higher cluster count in the unpolarized dataset truly indicates greater insect diversity, we analyzed the time/range community structure. Our hypothesis was that higher species specificity would correspond to a richer time/range community structure. However, our findings revealed that the HCA-derived community structure was weaker, particularly in the time dimension (**Fig 5**). This suggests that HCA's additional clusters may not correspond to distinct insect species but rather to over-sensitivity to variations in power spectra.

This over-sensitivity likely arises from the inherent differences in how HCA and GMM generalize power spectra patterns. HCA, being sensitive to variations in the relative powers of frequency peaks (55), may focus on differences between the powers of the fundamental frequency and its overtones. These differences can be due to varying observation aspects and could be accentuated for the fundamental peaks and a few harmonic overtones after averaging co-polarized and de-polarized signals.

In contrast, the GMM approach was applied not to the power spectra directly but to their UMAP-reduced representations. This transformed the 81-dimensional power spectra into a three-dimensional representation that aims to preserve the global structure and relationships between observations rather than focusing on specific frequencies and powers. Consequently, this makes GMM clustering less prone to overfitting and reduces sensitivity to individual spectral components.

Despite observing different granularity at which datasets are partitioned, we argue that the total cluster count remains a valid proxy for diversity, provided that the same approach is consistently used in comparative studies and reliably scales the number of clusters with actual insect diversity. This has been demonstrated in previous research using photonic sensors coupled with HCA clustering (9). Therefore, to evaluate the performance of polarimetric lidar, we shift our focus from analyzing cluster count to analyzing clusters' polarimetric properties.

Our comparative analysis of clusters retrieved from unpolarized and DoLP datasets reveals that the unpolarized approach struggles to co-cluster observations with low DoLP values. However, its clusters exhibit significant DoLP differentiation from random ones within the glossiest ($Q_1$) and most diffuse ($Q_4$) DoLP quartiles (compare **Figs 3** and **S4**). Moreover, incorporating polarimetric information only minimally rearranges observations (~10%) across non-adjacent DoLP quartiles (**Fig 4**). This suggests that unpolarized backscatter retains sufficient information on wing glossiness to effectively co-cluster the majority of DoLP-similar observations.

Furthermore, our comparison of results from co-polarized and DoLP datasets indicates that they yield similar diversity estimates. Also, both HCA and GMM produce DoLP-homogeneous clusters (**Fig 3** and **S4-5 Figs**), with the strongest agreement observed within the top glossy clusters ($Q_1$ group) (**Fig 4**). This suggests that most information on wing glossiness is derived from the harmonic content of co-polarized power spectra, while DoLP quantification remains valuable for identifying rare low-DoLP cases.

Our findings underscore the interplay between instrument sensitivity to insect morphology and the chosen clustering methodology. We find that while polarimetric lidar provides additional information, much of the relevant information is also present in unpolarized data, suggesting a need to balance instrument complexity with research goals. Furthermore, our findings highlight the importance of understanding the biases inherent to different clustering algorithms, as these can significantly influence diversity estimates.

# Methods

## HCA

We conducted Hierarchical Cluster Analysis (HCA) on area-normalized, log-transformed power spectra using MATLAB's **linkage** function, with **'ward'** specified as the method and **'euclidean'** as the metric. This method employs Euclidean distance to cluster power spectra based on similarity, accommodating minor variations in Wing Beat Frequencies (WBFs), a phenomenon frequently observed within the same species (9). Furthermore, this metric is sensitive to changes in the Degree of Linear Polarization (DoLP), including variations in the number of harmonic overtones and how power spectra scale with DoLP. We selected Ward's linkage criterion (56) to minimize the variance within newly formed clusters, thereby ensuring that observations within each cluster closely resemble the cluster's centroid.

To determine the optimal number of clusters, we analyze the changes in linkage rates, identifying significant deviations from the expected values due to random variations in power spectra. **Fig 9** illustrates our method. Panel **a** presents the linkage values in reverse order (from largest to smallest). By displaying these values on a logarithmic scale, we linearize the decrease in linkage values. From this plot, we calculate the linkage rates (slopes) at each step of the HCA and determine the median slope ($\gamma = -0.357$), which is depicted in **Fig 9a** as a solid line. This slope represents the expected decrease in linkage under conditions of random spectral variation.

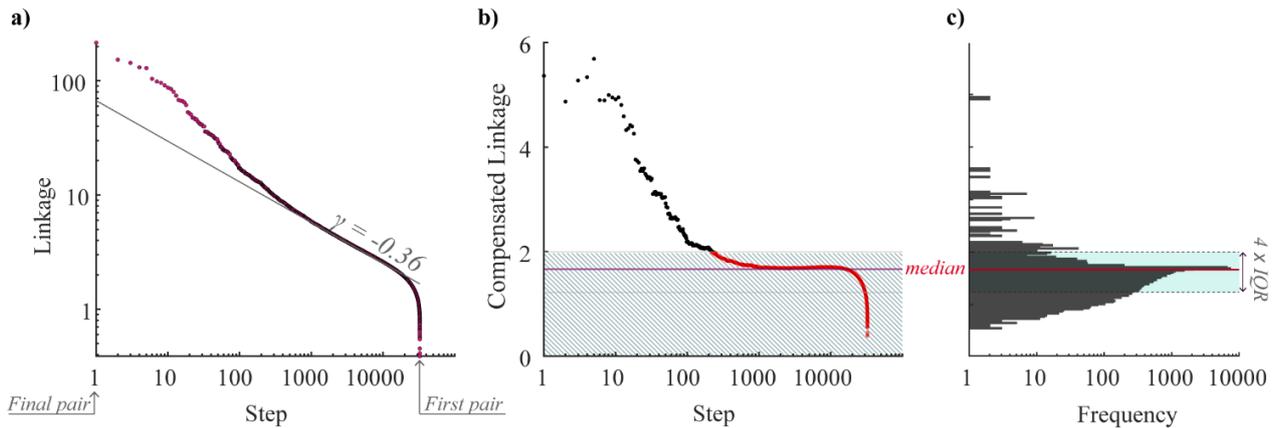

**Fig 9. Identifying optimal cluster numbers in hierarchical cluster analysis.** (a) Reverse-ordered linkage values on a logarithmic scale. The median slope (γ, solid line) represents the expected linkage decrease. (b) Compensated linkage values. Shaded area highlights the expected linkage. (c) Distribution of compensated linkage values with median (red line) and outlier boundaries ($Q_1 - 1.5 \cdot IQR$, $Q_3 + 1.5 \cdot IQR$, blue shaded area).

Next, to identify significant linkages, we calculate compensated linkage values using the formula $L_i^* = (i/N)^\gamma \cdot L_i$, where $L_i^*$ represents the compensated linkage, $L_i$ is the reversed linkage (from largest to smallest), $\gamma$ is the median slope, and $i$ ranges from 1 to the total number of steps, $N$. This transformation effectively modifies the linkage plot from **Fig 9a** to **Fig 9b**. Subsequently, we analyze the distribution of these compensated linkage values (**Fig 9c**) and identify significant linkages (outliers) using the 1.5xIQR rule (57). Specifically, we select those linkage values that exceed $Q_3 + 1.5 \cdot IQR$. The optimal number of clusters is then determined by the count of these outliers, as illustrated above the shaded area in **Fig 9b**.

## GMM

Prior to clustering lidar observations using GMM, we reduced the dimensionality of area-normalized, log-transformed power spectra using Uniform Manifold Approximation and Projection (UMAP) (58), MATLAB implementation (59). UMAP parameters were **n_components = 3, dmin = 0.01, n_neighbours = 199**, and **metric=**

'euclidean'. The **dmin** parameter is chosen to achieve tighter grouping of similar observations, while **n_neighbours** balanced algorithm between focusing on local and global structure of the data. We chose the maximal **n_neighbours** value allowed by the UMAP library. Reducing the data from 81 features (frequencies) to three (UMAP-coordinates) increased data point density, aiding a density-based GMM algorithm to identify clusters.

Next, we fit a Gaussian mixture distribution (60) to the UMAP-embedded data using MATLAB's **fitgmdist** function (Statistics and Machine Learning Toolbox). To determine the optimal number of clusters, we scanned the **n_components** parameter (range: 55 – 555) and selected the solution minimizing the Bayesian Information Criterion (BIC). BIC is calculated as $BIC = ln(n)k - 2ln(L)$, where $n$ is the number of observations, $k$ is the number of estimated parameters, and $L$ is the maximum value of the likelihood function for the model. Other **fitgmdist** parameters were: **RegularizationValue = 1e-6, CovarianceType = 'full', SharedCovariance = 'false', Replicates = 1**, and **Options= statset (MaxIter = 100, TolFun = 1e-3)**.

Another approach to finding the optimal number of clusters would be to use Akaike Information Criterion (AIC) calculated as $AIC = 2k - 2ln(L)$. Both AIC and BIC criteria favor models that fit data well (large $L$) and have fewer parameters (small $k$), however BIC tends to impose a stronger penalty on the number of parameters, resulting in favoring simpler models than AIC.

## Evaluating clustering agreement

Next, we evaluate how well clustering algorithms agree about the optimal partitioning of the data. To compare solutions, we leverage two metrics from the scikit-learn library in Python (61): Adjusted Mutual Information Score (AMI) and Homogeneity Score. AMI (62) is a variation of Mutual Information (MI) that accounts for a chance for two solutions to agree, especially when we compare clusterings of different sizes or with different numbers of clusters. AMI scores range from 0 to 1, with 1 indicating perfect agreement and score of 0 indicating agreement no better than random chance.

We also employ a Homogeneity score (63), a metric that reflects the internal consistency of solutions, for example, if larger clusters in one solution are split into many in another. A homogeneity score of 0 indicates that clusters of one solution have random observations compared to another solution. A score of 1 indicates perfect homogeneity, with each cluster in one solution containing observations of the same cluster in another.

## Time and range communities

We analyze the time and range profiles associated with clusters by extracting the time and range stamps of assigned observations. For each cluster, we calculate the probability of observing a member at specific time and range bins (as in **Figs 1f** and **1g**). To compare clusters' time and range distributions, we employ the two-sample Kolmogorov-Smirnov (K-S) test (64), implemented in MATLAB, Statistics and Machine Learning Toolbox. The K-S test assesses whether two empirical distributions originated from the same parent distribution, providing a distance metric and p-value. Using K-S p-values measured between cluster pairs, we construct two similarity matrices: one for time and another for range.

We construct similarity matrices to understand how clusters naturally group into communities. These communities are characterized by greater internal similarity compared to their similarity with clusters outside the group. To identify these communities, we first calculate a modularity matrix (65) using **modularity_f(A, gamma)**, implemented in an external MATLAB library (66), where:

1. **A:** The similarity matrix (K-S p-values) that contains K-S p-values between all cluster pairs.
2. **gamma (γ):** The resolution parameter controlling the granularity of the community structure. Lower values (γ < 1) tend to produce fewer communities, while higher values (γ > 1) result in more communities. In our analysis, we use the default value of γ = 1.

GenLouvain algorithm (66), with deterministic output and default parameters, is then applied to the modularity matrix. This yields a community assignment for each cluster, effectively partitioning the clusters into time and range communities.

Additionally, based on this cluster-to-community mapping, we calculate a modularity score ($M$), which quantifies the strength of the identified community structure. Modularity values range from 0 (indicating a random structure) to 1 (signifying a well-defined structure), or even -1 (suggesting a structure less optimal than random).

## Lidar based diversity indices

To quantify and compare clustering results across experiments, we employed Hill numbers (67), a family of diversity metrics that allows us to emphasize different aspects of diversity by adjusting a single parameter, $\alpha$ (68). Hill numbers are expressed by the following equation (**Eq. 4**):

$$H_\alpha = \left[ \sum_{j=1}^{S} p_j^\alpha \right]^{\frac{1}{1-\alpha}} \tag{4}$$

, where $S$ is a set of all clusters, $p_j$ is a relative size of cluster $j$ (cluster $j \in S$) calculated as the number of observations in cluster $j$ divided by the total observations, and $\alpha$ is an integer value ranging from $\pm\infty$.

Varying $\alpha$, we land at three diversity indices:

**Total number of clusters.** The $H_0$ metric ($\alpha = 0$) reflects the total number of clusters (species) $S$, giving a high importance to rare clusters (**eq. 5**):

$$H_0 = \sum_{i=1}^{S} p_i^0 = S \tag{5}$$

**Effective number of clusters.** The $H_1$ ($\alpha = 1$), also known as Shannon diversity of order 1, weighs both rare and abundant clusters (69), providing an estimate of how many equally-sized clusters would yield the same Shannon Entropy (**Eq. 6, 7**). This is analogous to the number of effective choices in a prediction model.

$$H_1 = exp(H') \tag{6}$$

$$H' = -\sum_{i=1}^{S} p_i ln(p_i) \tag{7}$$

**The number of dominant clusters.** The $H_2$ metric ($\alpha = 2$) emphasizes dominant clusters, indicating a more even spread of diversity across clusters (**Eq. 8**).

$$H_2 = 1/\sum_{i=1}^{S} p_i^2 \tag{8}$$

## Detrending of power spectra

For visualization purposes, we detrended the power spectra by fitting a line (trend) to area-normalized and log-transformed power spectra and then subtracting it. The resulting positive and negative values indicate power above and below the trend. This approach, applied for heatmap visualization with a diverging colormap, allows us to highlight even subtle oscillations. However, it is important to note that we do not use the detrended power spectra in any analysis.

# Bootstrapping to evaluate confidence intervals

To assess the variability of our metrics, we employed a bootstrapping technique (70), a resampling-based method well-suited for scenarios with limited day-to-day data. This approach involves generating N = 1000 synthetic datasets by randomly sampling observations with replacement from the original dataset. Each original observation has an equal probability of being included in a synthetic dataset, and some may be included multiple times. The original dataset can be observations from the same cluster, community, or any other relevant subset.

For each synthetic sample, we calculate the metric of interest, resulting in N = 1000 variants depending on the drawn observations. From this distribution, we empirically estimate the mean of the metric and its 95% confidence intervals (CIs) using the $2.5^{th}$ and $97.5^{th}$ percentiles. This provides a range within which we are 95% confident that the true value of the metric lies, accounting for sampling variability.

In our study, we applied bootstrapping to estimate confidence intervals (CIs) for several key metrics:

1. **Clusters' mean DoLP:** To assess the DoLP for both found and random clusters, we generated N = 1000 synthetic samples for each cluster by randomly drawing observations with replacement from the evaluated cluster. For each synthetic sample, we calculated the mean DoLP. By retrieving N values of mean DoLP, we then evaluated this distribution to obtain the mean and CIs for the cluster's DoLP.
2. **Time and Range Profiles:** For each time/range community or range-DoLP subset, we generated N = 1000 synthetic samples by randomly drawing observations with replacement. For each synthetic sample, we determined the probability of an observation in time (or range). To quantify the variability of these probabilities, we analyzed the N values obtained at each time (or range) bin, reporting the mean probability and its 95% CIs.

# Acknowledgments

The lidar instrumentation were in part, kindly provided by Norsk Elektro Optikk A/S, Norway. We thank Ebba von Wachenfeldt, Zachary Nolen, Magne Friberg, Jadranka Rota and in particular, Jens Rydell for assistance in field work, may he rest in peace. We thank Rachel Muheim for receiving us at the Stensoffa ecological field station. We thank Zhicheng Xu and Jacobo Salvador for discussion and initial data analysis.

# Supporting information

**S1 Fig. AMI and homogeneity scores between HCA and GMM clusters.**

**S2 Fig. 35 largest clusters (HCA, DoLP dataset)**

**S3 Fig. 35 largest clusters (GMM, DoLP dataset)**

**S1 Text. Diversity indices variability due to random UMAP/GMM initialization.**

**S4 Fig. DoLP characterization of clustering results (HCA, un-pol. and co-pol. datasets).** Comparison of HCA clustering results (black) with random clustering (gray).

**S5 Fig. DoLP characterization of clustering results (GMM, un-pol. and co-pol. datasets).** Comparison of DoLP clustering results (black) with random clustering (gray).

**S6 Fig. Median power spectra of clusters that belong to various DoLP ranks (HCA, DoLP dataset).**

**S7 Fig. Median power spectra of clusters that belong to various DoLP ranks (GMM, DoLP dataset**).

**S8 Fig. Characterization of time and range communities.** Probability distributions for range (**ABC**) and time (**I-II-III**) communities. **Heatmaps at the ABC and I-II-III intersection** display median power spectra for each time-range community.

**S9 Fig. Range dependence of co-polarized backscatter.** Probability distributions show the likelihood of observations within range communities (A, B, C) and DoLP quartiles (Q1-Q4), with heatmaps of corresponding power spectra. Note the probability spike in C-plots (red dot) co-occurred with the land piece left of the laser beam over the pond.

**S1 Fig. AMI and homogeneity scores between HCA and GMM clusters**

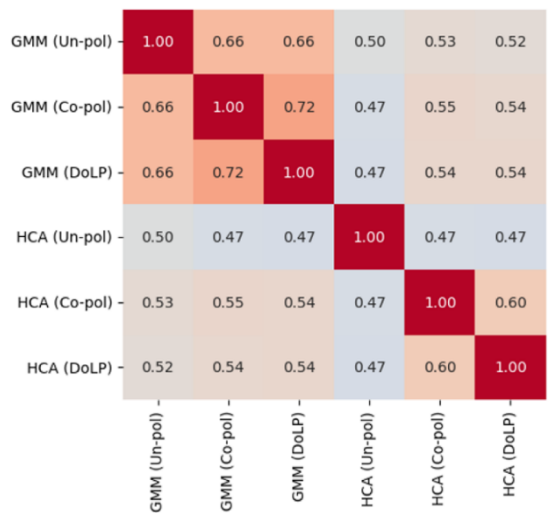
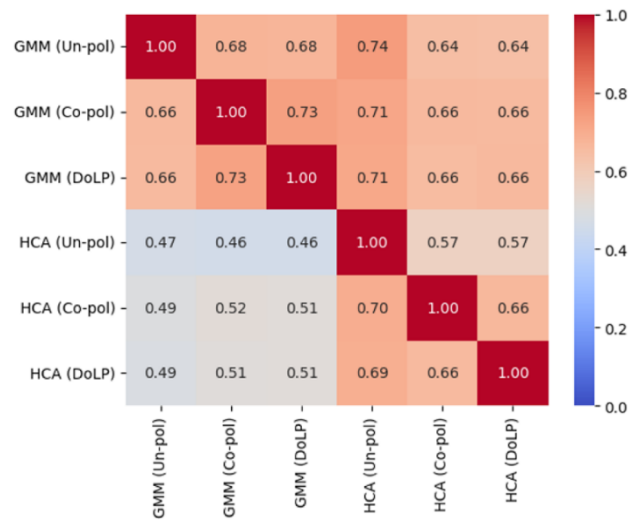

**S1 Fig. AMI and homogeneity scores between HCA and GMM clusters.**

**S2 Fig. 35 largest clusters (HCA, DoLP dataset)**

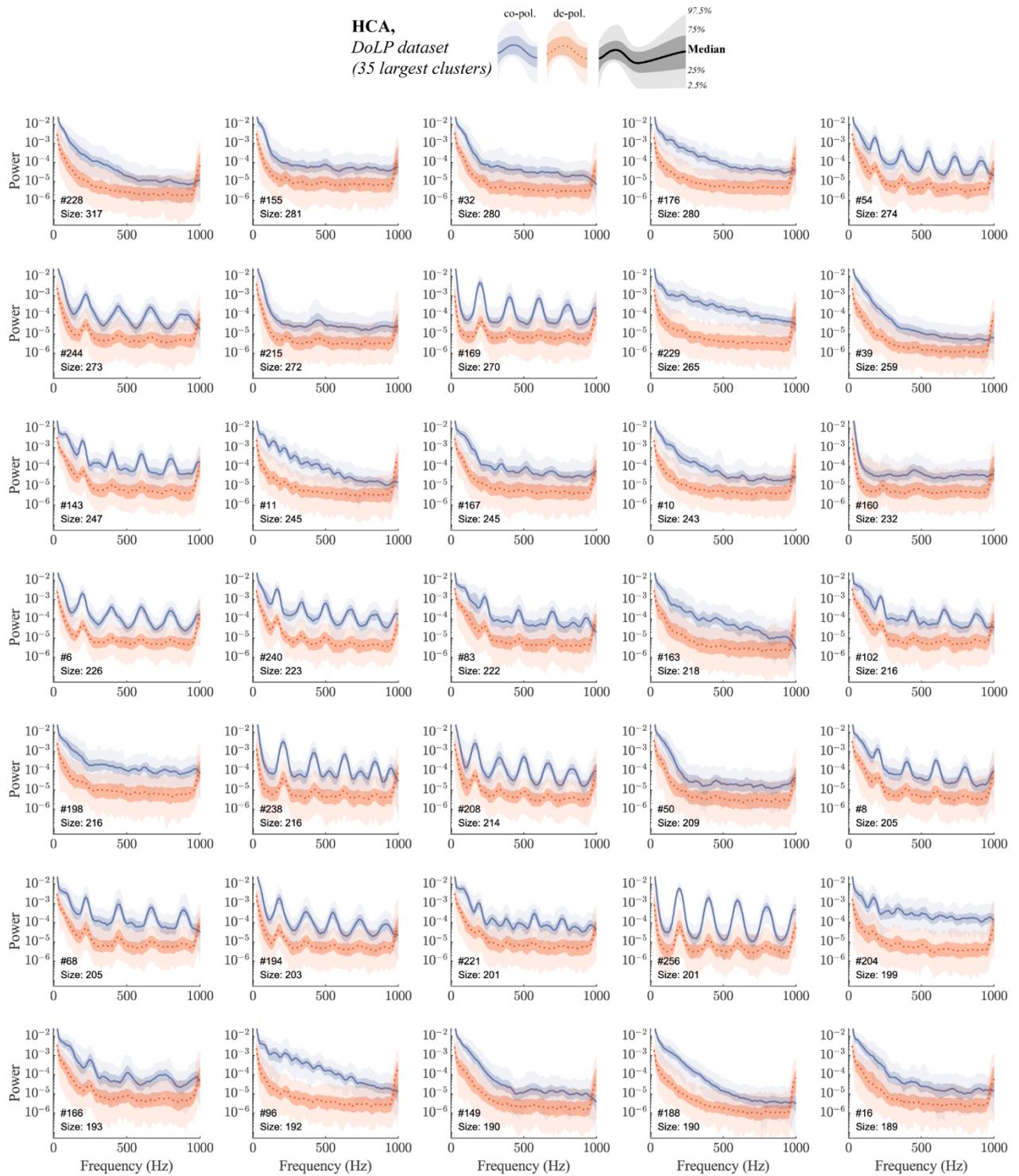

# S3 Fig. 35 largest clusters (GMM, DoLP dataset)

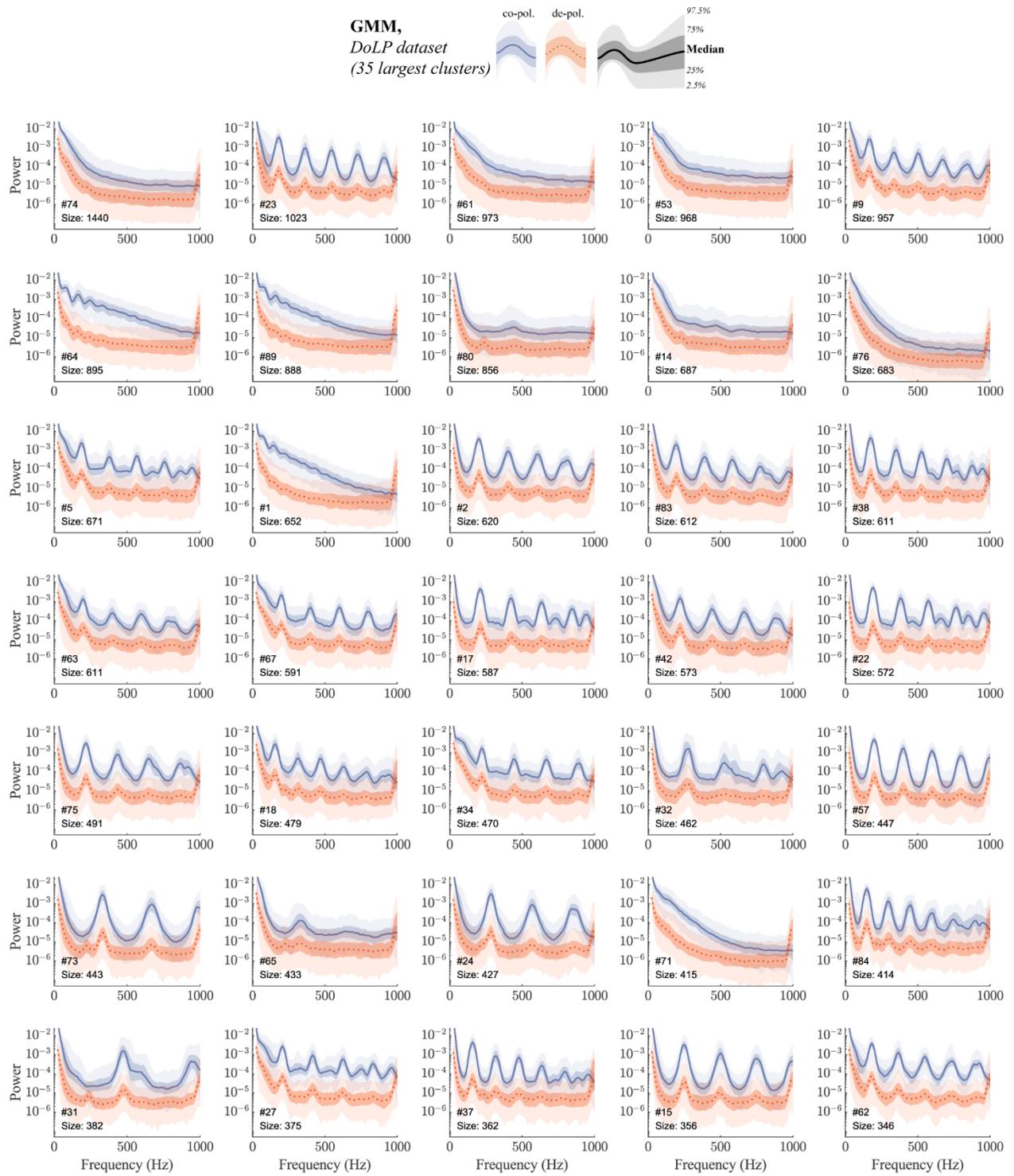

**S1 Text. Diversity indices variability due to random UMAP/GMM initialization**

We also investigated the impact of random initialization of the UMAP and GMM algorithms on clustering results. Given they are stochastic, each algorithms' execution can yield different labeling solutions, potentially affecting the consistency of derived diversity indices.

To quantify variability of diversity indices, we perform 100 runs of UMAP embedding followed by GMM clustering. Each run, we change the random seed for both algorithms, while maintaining other parameters as described in Section **Methods: GMM**. Specifically, we keep the same value for the number of components, as was found to be optimal when using BIC score (see **S1 Table**). We summarize the results of these 100 UMAP/GMM runs in **S2 Table**, presenting the mean diversity indices in bold, along with their corresponding confidence intervals in italics (2.5$^{th}$ and 97.5$^{th}$ percentiles).

**S1 Table. Optimal solutions from UMAP and GMM algorithms initialized with 'random seed' = 42 (result reported in the main text).**

|  | Dataset | Number of components (GMM model) | Number of clusters found, $NoC$, ($H_0$) | BIC (for optimal solution) |
|---|---|---|---|---|
| GMM | un-pol | 80 | 80 | 2.091e+05 |
|  | co-pol | 87 | 86 | 2.040e+05 |
|  | DoLP | 89 | 89 | 2.020e+05 |

**S2 Table. Variations in diversity indices resulting from 100 random initializations of UMAP and GMM.**

|  | Dataset | $H_0, N_{cl}$ | $H'$ | $H_1$ | $H_2$ | BIC |
|---|---|---|---|---|---|---|
| GMM | un-pol | **79.34**<br>*77...80* | **4.13**<br>*4.05...4.19* | **62**<br>*57.17...66.02* | **54.25**<br>*47.85...59.22* | **2.108e+05**<br>*2.082e+05...2.136e+05* |
|  | co-pol | **85.44**<br>*82...87* | **4.18**<br>*4.10...4.24* | **65.16**<br>*60.14...69.29* | **56.51**<br>*50.24...61.63* | **2.059e+05**<br>*2.034+05...2.088e+05* |
|  | DoLP | **87.34**<br>*84...89* | **4.19**<br>*4.11...4.27* | **66.02**<br>*60.81...71.51* | **56.43**<br>*51.22...62.21* | **2.050e+05**<br>*2.023e+05...2.081e+05* |

We observed that the number of found clusters ($H_0$) varied by up to ±2 across all three datasets, with significantly fewer clusters found in the unpolarized dataset compared to the co-polarized and DoLP datasets. Despite these minor fluctuations in $H_0$, random initialization introduced variability of approximately ±5 cluster for the effective ($H_1$) and dominant ($H_2$) cluster numbers. These results show that the unpolarized and DoLP datasets differ significantly in the number of clusters ($H_0$), and therefore, the DoLP dataset shows a higher richness of signal. However, the variability of other indices is too high to confidently determine if these diversity estimates are significantly different between datasets.

By considering both optimal and suboptimal clustering solutions (as indicated by the BIC variability in **S2 Table**), this analysis provided insights into the stability and robustness of diversity indices in the presence of stochastic algorithms. Furthermore, it allowed us to assess whether the three datasets exhibited distinct diversity profiles, irrespective of the specific clustering solution obtained in each run.

**S4 Fig. DoLP characterization of clustering results (HCA, unpol. and co-pol. datasets).**

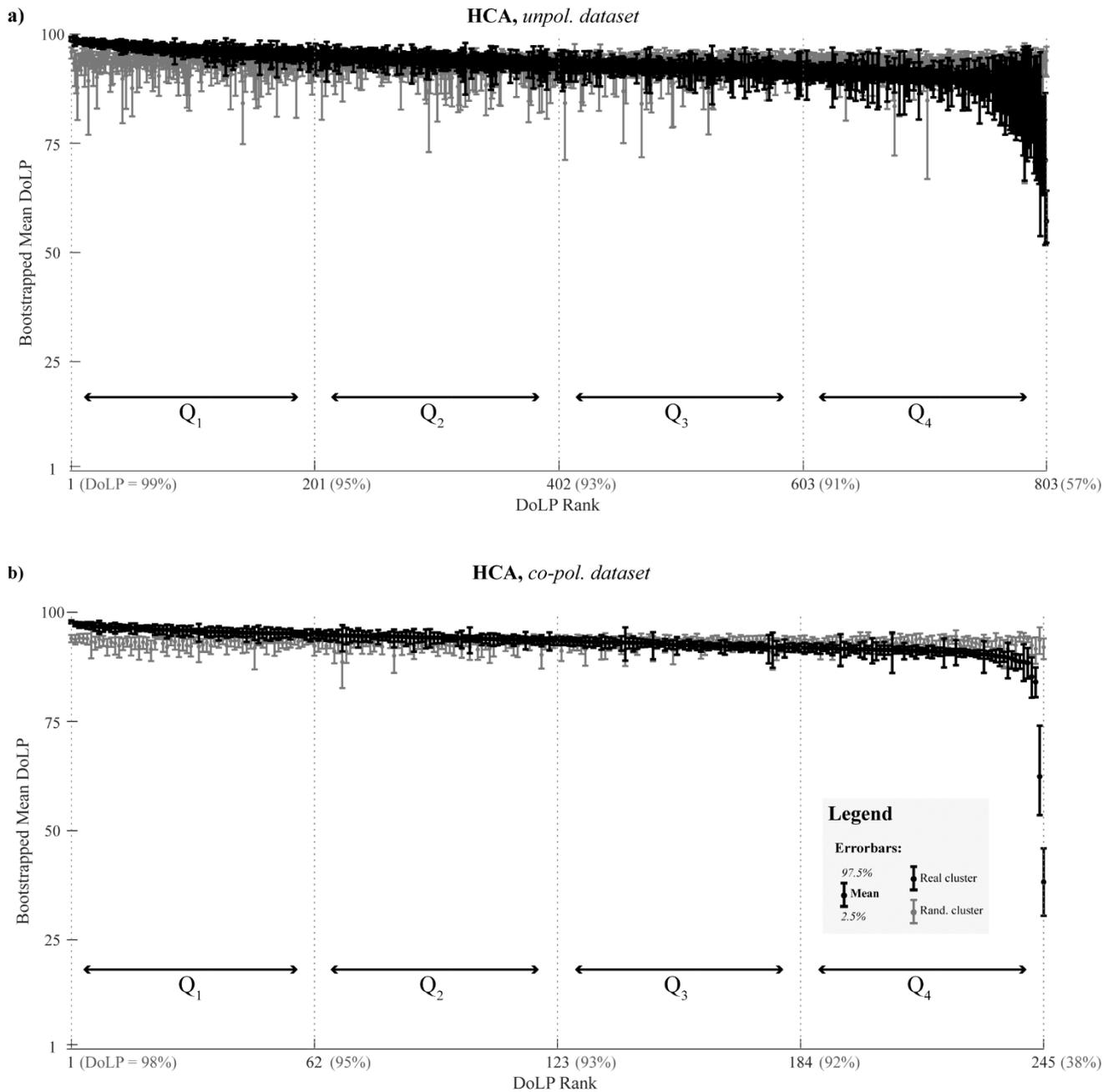

**S4 Fig. DoLP characterization of clustering results (HCA, unpol. and co-pol. datasets).** Comparison of HCA clustering results (black) with random clustering (gray).

**S5 Fig. DoLP characterization of clustering results (GMM, unpol. and co-pol. datasets).**

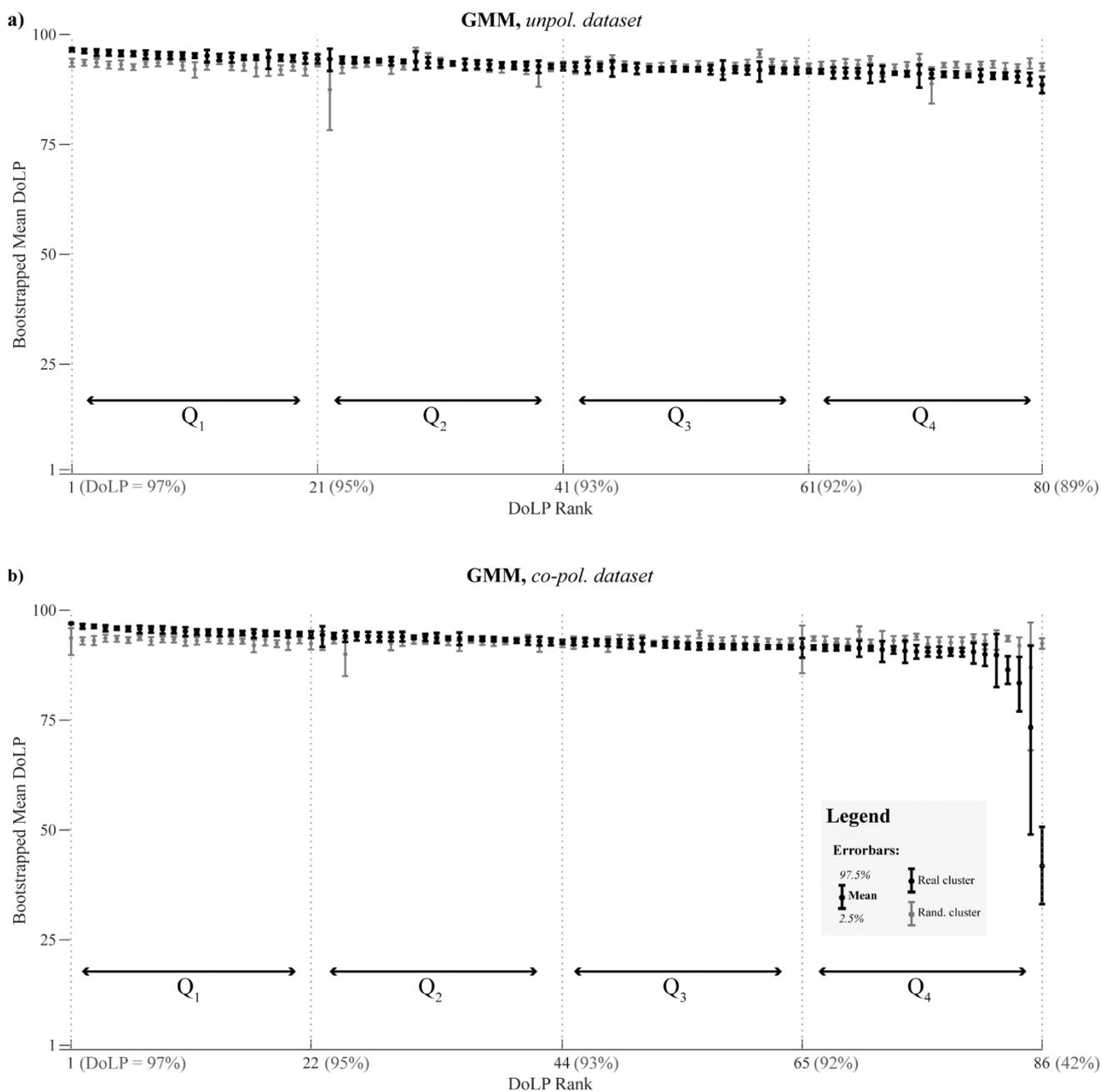

**S5 Fig. DoLP characterization of clustering results (GMM, unpol. and co-pol. datasets)**. Comparison of GMM clustering results (black) with random clustering (gray).

**S6 Fig. Median power spectra of clusters that belong to various DoLP ranks (HCA, DoLP dataset).**

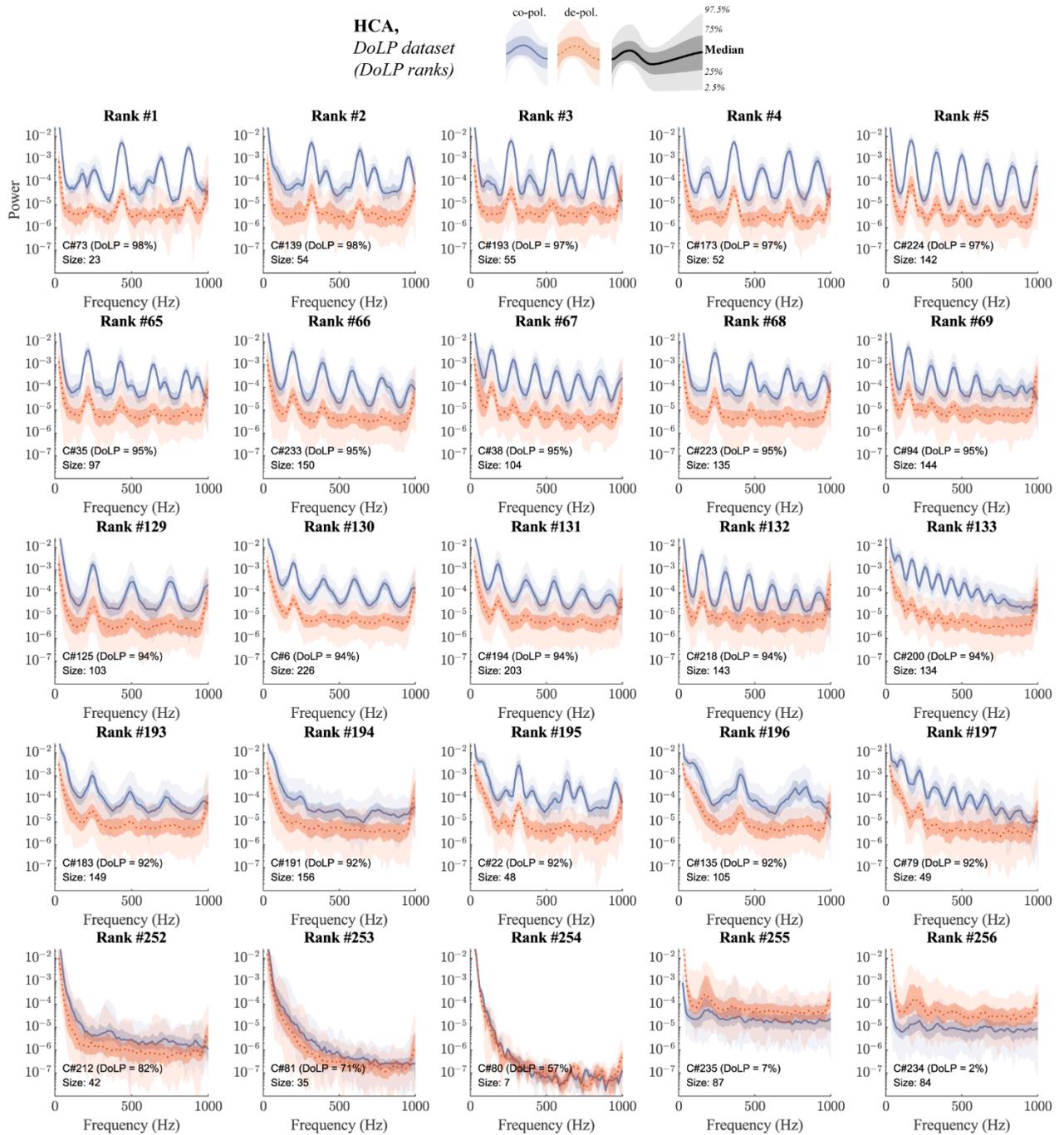

S6 Fig. Median power spectra of clusters that belong to various DoLP ranks (HCA, DoLP dataset).

**S7 Fig. Median power spectra of clusters that belong to various DoLP ranks (GMM, DoLP dataset).**

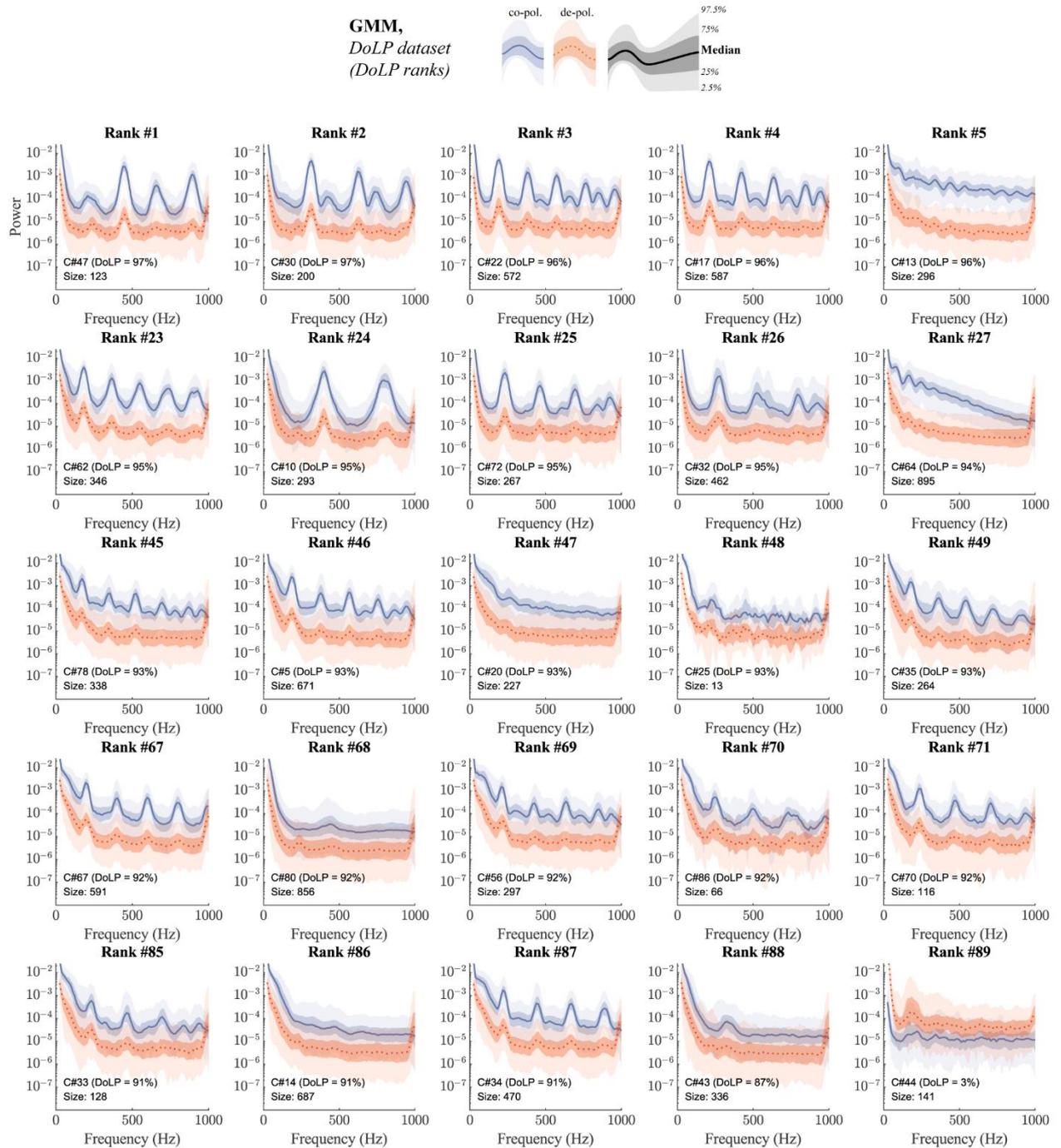

**S7 Fig. Median power spectra of clusters that belong to various DoLP ranks (GMM, DoLP dataset).**

**S8 Fig. Characterization of time and range communities (GMM, DoLP dataset)**

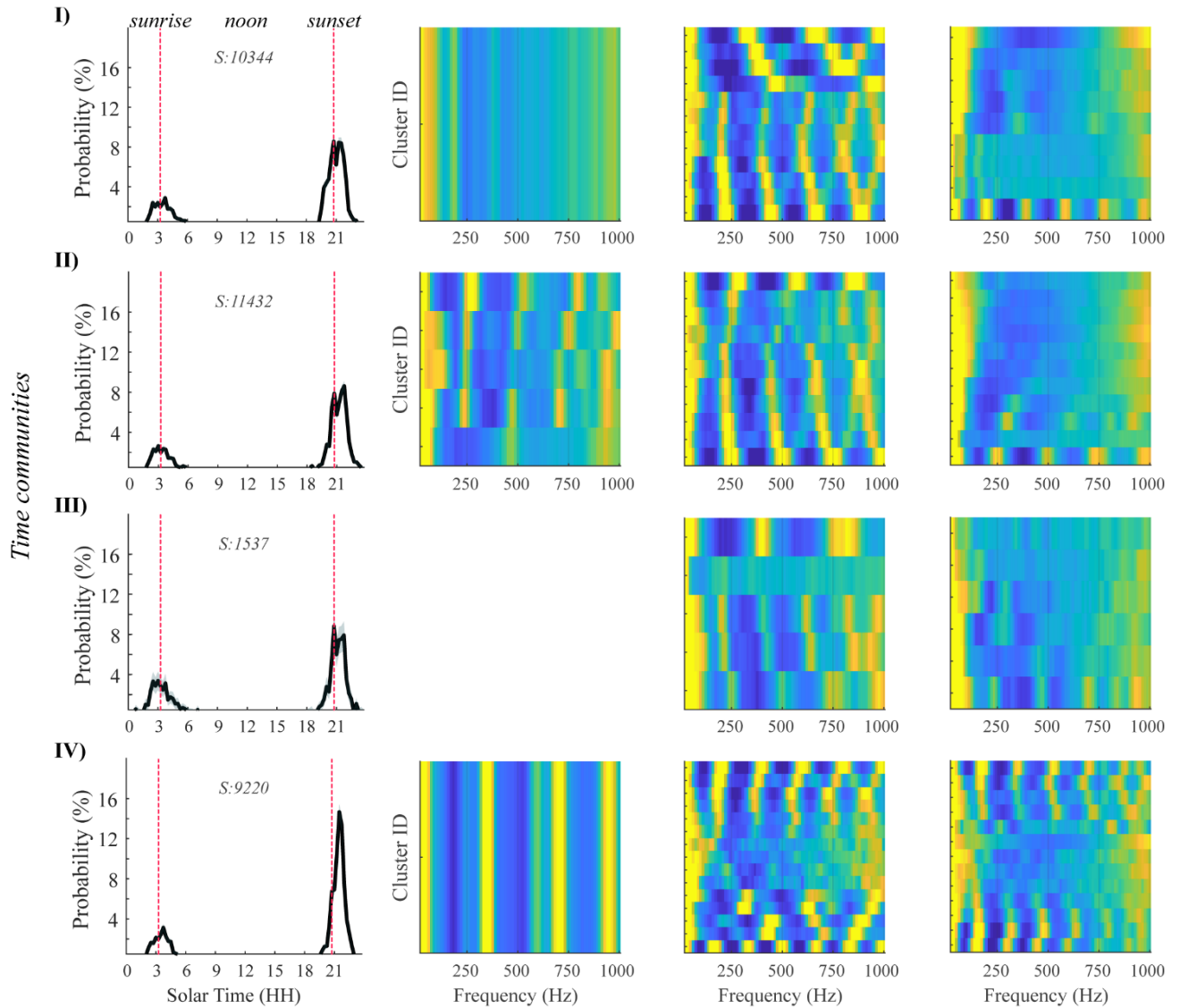

**S8 Fig. Characterization of time and range communities.** Probability distributions for range (**ABC**) and time (**I-II-III**) communities. **Heatmaps at the ABC and I-II-III intersection** display median power spectra for each time-range community.

**S9 Fig. Range dependence of co-polarized backscatter (GMM, DoLP dataset)**

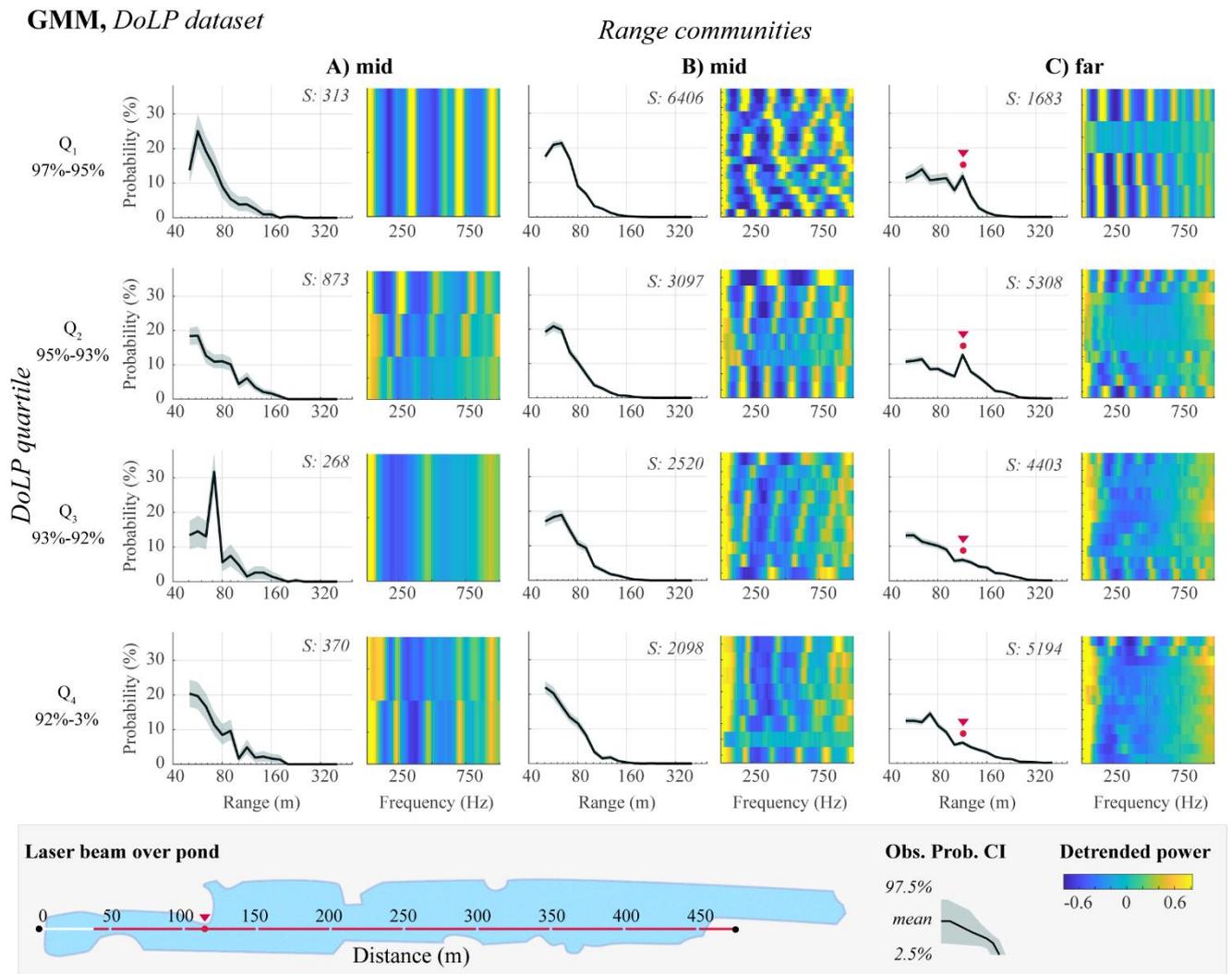

**S9 Fig. Range dependence of co-polarized backscatter.** Probability distributions show the likelihood of observations within range communities (A, B, C) and DoLP quartiles ($Q_1$-$Q_4$), with heatmaps of corresponding power spectra. Note the probability spike in C-plots (red dot) co-occurred with the land piece left of the laser beam over the pond.